\documentclass[twocolumn,secnumarabic,amssymb,nobibnotes,aps,pra,superscriptaddress]{revtex4-1}
\usepackage{
  color,
  graphicx,
  amsfonts,
  amsmath,
  float,
  physics
}
\usepackage[utf8]{inputenc}
\usepackage{tikz}

\definecolor{c11}{cmyk}{1,0,0,0}
\definecolor{c12}{cmyk}{0.73,0.76,0,0}
\definecolor{c21}{cmyk}{0.26,0.80,0,0}
\definecolor{c22}{cmyk}{0,0.95,0.92,0}
\definecolor{cat}{cmyk}{0.72,0,1,0}
\DeclareMathOperator{\e}{e}

\renewcommand{\vec}[1]{\ensuremath{\mathbf{#1}}}
\graphicspath{{./gnuplot/}}

\begin{document}

\title{Dynamical distortions of structural signatures in molecular High Harmonic Spectroscopy}

\author{Marie Labeye}
\author{Fran\c cois Risoud}
\affiliation{Sorbonne Université, CNRS, Laboratoire de Chimie Physique-Mati\`ere et Rayonnement, 75005 Paris, France}
\author{Camille L\'ev\^eque}
\altaffiliation[Present address:]{Wolfgang Pauli Institute c/o Faculty of Mathematics, University of Vienna, Oskar-Morgenstern Platz 1, A-1090 Vienna, Austria \& Vienna Center for Quantum Science and Technology, Atominstitut, TU Wien, Stadionallee 2, A-1020 Vienna, Austria
}
\affiliation{Sorbonne Université, CNRS, Laboratoire de Chimie Physique-Mati\`ere et Rayonnement, 75005 Paris, France}
\affiliation{Theoretische Chemie, Physikalisch-Chemisches Institut, Universität Heidelberg, Im Neuenheimer Feld 229, D-69120 Heidelberg, Germany}
\author{J\'er\'emie Caillat}
\author{Alfred Maquet}
\affiliation{Sorbonne Université, CNRS, Laboratoire de Chimie Physique-Mati\`ere et Rayonnement, 75005 Paris, France}
\author{Tahir Shaaran}
\author{Pascal Sali\`eres}
\affiliation{LIDYL, CEA, CNRS, Université Paris-Saclay, CEA-Saclay, 91191 Gif-sur-Yvette, France}
\author{Richard Ta\"ieb}
\affiliation{Sorbonne Université, CNRS, Laboratoire de Chimie Physique-Mati\`ere et Rayonnement, 75005 Paris, France}

\begin{abstract}
We study the signature of two-center interferences in molecular high-order harmonic spectra, with an emphasis on the spectral phase. With the help of both ab initio computations based on the time-dependent Schr\"odinger equation and the molecular Strong-Field Approximation (SFA) as developed by Chirilă et al. [Physical Review A, 73, 023410 (2006)] and Faria [Physical Review A, 76, 043407 (2007)], we observe that the phase behavior is radically different for the short and the long trajectory contributions. By means of Taylor expansions of the molecular SFA, we link this effect to the dynamics of the electron in the continuum. More precisely, we find that the value of the electric field at recombination time plays a crucial role on the shape of the destructive interference phase-jump.
\end{abstract}

\pacs{33.80.Rv, 42.65.Ky}

\maketitle

\section{Introduction}

Over the last decades, High-order Harmonic Generation (HHG) has raised an increasing interest in the field of strong laser physics with the development of attosecond science \cite{scrinzi_attosecond_2006,krausz_attosecond_2009}.
Two main classes of experiments have emerged in this context. First, HHG is a robust tool to generate single or trains of attosecond pulses of coherent XUV radiation \cite{farkas_proposal_1992,antoine_attosecond_1996,antoine_generation_1997,christov_generation_1998,paul_observation_2001,mairesse_attosecond_2003,dudovich_measuring_2006,gaarde_macroscopic_2008}, that can be used to perform electronic or nuclear pump-probe spectroscopy on the attosecond scale with \AA ngstr\"om resolution \cite{cavalieri_attosecond_2007,haessler_phase-resolved_2009,varju_physics_2009,klunder_probing_2011,caillat_attosecond_2011,leveque_direct_2014}. Second, HHG can be  employed as a self-probing tool to monitor attosecond dynamics in the generating medium \cite{lein_attosecond_2005,uiberacker_attosecond_2007,baker_probing_2006,baker_dynamic_2008,ramakrishna_high-order_2008,haessler_attosecond_2010,zhao_positioning_2012}, e.g.  to implement tomography of molecular orbitals \cite{itatani_tomographic_2004,vozzi_generalized_2011,salieres_imaging_2012}. References \cite{haessler_self-probing_2011, CDLin2010} provide a comprehensive review on the self-probing applications of HHG.

Most of these studies showed that the intensity and the phase of the generated harmonics encode structural and dynamical information. While the measurement of their intensity, i.e. counting the photon number for each wavelength, is rather straightforward, obtaining their phases was more difficult. Nowadays, several experimental techniques based on interferometry are mastered to access the harmonic phase, such as RABBIT (Reconstruction of Attosecond Beating by Interference of Two-photon Transition) \cite{veniard_phase_1996,muller_reconstruction_2014}, streaking \cite{constant_methods_1997,itatani_attosecond_2002}, TSOIN (Two-Source Optical INterferometry) \cite{smirnova_high_2009,zhou_molecular_2008,camper_high-harmonic_2014},  PROOF (Phase Retrieval by Omega Oscillation Filtering) \cite{chini_characterizing_2010}, MAMMOTH (Mixed Approaches for the MeasureMent Of the Total Harmonic phases) and CHASSEUR (Combined Harmonic Attosecond Spectroscopy by two-Source EUv interferometry and Rabbit) \cite{camper_spectroscopie_2014}, or other approaches \cite{mcfarland_high-order_2009,mairesse_phase_2010}.
These measures of the harmonic phase are crucial to study e.g. the dynamics of auto-ionizing states or Fano resonances \cite{le_extraction_2008,haessler_phase-resolved_2009,strelkov_role_2010,strelkov_high-order_2014,rothhardt_enhancing_2014,camper_spectroscopie_2014, Kotur_2016, Gruson_2016, Kaldun_2016, Argenti_2017}, interferences between different ionization paths \cite{smirnova_high_2009,mairesse_high_2010,figueira_de_morisson_faria_molecular_2010,diveki_spectrally_2012} and two-center interferences \cite{wagner_extracting_2007,ciappina_influence_2007,boutu_coherent_2008,zhou_molecular_2008,chirila_explanation_2009,etches_two-center_2011,spiewanowski_high-order-harmonic_2013,spiewanowski_field-induced_2014}.

To picture the phenomenon of HHG, the well known three-step model is invoked  \cite{corkum_plasma_1993,schafer_above_1993}: (i) under the influence of a strong laser field, an electron is first pulled out from the atom or molecule, then (ii) scattered in the continuum and driven back to the parent ion, where (iii) it finally may recombine into its initial state,  releasing radiatively its accumulated kinetic energy. The emitted photons constitute the harmonic spectrum. In this model the electron in the continuum is considered as a classical particle and thus its motion is characterized by \textit{trajectories}. Within the Strong-Field Approximation (SFA) \cite{lewenstein_theory_1994}, which is a fully quantum theory, the notion of electron trajectories (or quantum orbits) in the continuum arises when searching saddle-point solutions in the equations leading to the emitted dipole \cite{salieres_feynmans_2001,lewenstein_phase_1995,milosevic_role_2002}, thus validating the three-step picture. Each harmonic of the spectrum is mainly emitted by two possible trajectories of the recolliding electron: the short and the long ones. Both lead to the same kinetic energy but with different times at which the electron recollides with the parent ion. The long trajectories correspond to electrons released before and recolliding after the short ones. Within a quantum theory, they interfere and shape the total spectrum with constructive and destructive patterns.

In the case of molecules, one may envision that electrons ionized from one atomic center and recombining to another will form additional classes of trajectories. Therefore the structure of the system is encoded through additional interferences in the spectrum, that are reminiscent of Young’s two-slit phenomenon in diatomics \cite{lein_role_2002,lein_interference_2002}.

In a standard SFA saddle-point approach \cite{zhou_alignment_2005,zhou_role_2005,kamta_three-dimensional_2005}, the structure of the molecule is usually introduced via the transition matrix elements in step 1 and 3, whereas the electron dynamics , i.e. step 2, remains atomic. Within this approach, destructive interferences are observed in the harmonic spectrum of diatomic molecules that correspond to the zeros of the recombination dipole matrix element. This results in a discontinuous phase-jump of exactly $\pm\pi$ rad  in the theoretical spectra.

However, this behavior does not compare well with experimental data \cite{wagner_extracting_2007,boutu_coherent_2008} nor with ab initio computations based on the exact resolution of the Time Dependent Schr\"odinger Equation (TDSE) \cite{lein_role_2002,lein_interference_2002,van_der_zwan_two-center_2010}. The phase-jumps at the destructive interferences are found far from being sharp: they are smoothed, cover several harmonics and span less than $\pi$ rad.


Ten years ago, a method accounting for the effect of the molecular structure on the continuum electron dynamics in SFA computations, called molecular SFA, has been proposed in \cite{chirila_strong-field_2006} and closely investigated in \cite{faria_high-order_2007}. It relies on the separation of the {\it dipole} into four terms, each one exhibiting an additional phase incorporated to the electronic action. The saddle-point approximation is carried out for each term and leads to four groups of possible electron trajectories (with their subsets of short and long paths). They correspond to the ionization from one center and the recollision to either the same center or the other one. To our knowledge the shape of the phase-jumps in such calculations has not been studied or addressed, except in Ref. \cite{etches_two-center_2011} where similar computations have been performed in the specific case of aligned polar CO molecules. Interestingly, the phase-jumps are smoothed, what we will confirm with our simulations.

Other models based on the computation of SFA with field distorted orbitals also display smoothed phase-jumps \cite{spiewanowski_high-order-harmonic_2013,spiewanowski_field-induced_2014}. These approaches rely on an adiabatic dressing of the ground state by the instantaneous electric field at recombination. This modifies the shape of the corresponding dipole, and thus the shape of the spectrum. An important feature of this approach is that, if the recombination occurs at a time when the electric field is zero, then the recombination dipole is no longer modified by the dressing, and one recovers a sharp minimum in the spectrum. 

We propose here a complete examination of the  two-center interferences in homo-nuclear diatomic molecular models, beyond the smoothing of the structural phase jumps.
The motivation of this study arose when investigating HHG amplitude and phase for diatomic molecules as a function of the internuclear distance within ab initio computations based on the TDSE. We observed very different behaviors between the phase-jumps when discriminating the short and the long trajectories,  in apparent relation with the electric field at the recombination. To get physical insight with an analytical theory we compared with the results of molecular SFA \cite{chirila_strong-field_2006,faria_high-order_2007} and found that they reproduce the ab initio observations. Finally, we performed additional developments of the molecular SFA based on Taylor expansions and fully explain the shape of the phase-jump in relation with the electric field at the recombination time.

This paper gives a full account of our results following our preliminary studies presented in \cite{risoud_laser-induced_2017}. It is organized as follows. In Sec. \ref{sec_methods} we introduce the models we used. First we expose the ab initio computations we performed on a 1D molecular model by solving the TDSE. Second we recall the basic equations of atomic SFA and molecular SFA and expose our further developments based on Taylor expansions of the molecular SFA. The ab initio computations will serve as the reference to compare with the SFA computations. In Sec. \ref{sec_results} we display the results of the ab initio and the molecular SFA computations in the case of a model H$_2$. In Sec. \ref{sec_interpretation} we provide explanations of the behavior of the phase-jumps with the examination of the Taylor expansions of the molecular SFA. We finally summarize our work in Sec. \ref{sec_conclusion} and we will draw a qualitative parallel between the field distorted orbitals computations \cite{spiewanowski_high-order-harmonic_2013,spiewanowski_field-induced_2014}  and our developments. In this paper, all the equations are expressed in atomic units (a.u.).


\section{Methods}\label{sec_methods}

\subsection{Ab initio computations}\label{subsec_methods_ab_initio}

We consider a diatomic molecule subjected to a low-frequency strong laser field with a 1D model within the single-active electron and fixed nuclei approximation \cite{1D}, by solving numerically the TDSE. The computational efficiency of these simulations allows us to perform an extensive study of the problem. This approach serves as an exact reference
illustrating the validity of our revisited molecular SFA and its interpretation presented hereafter.

We use a double-well potential $V(x)$ as in \cite{chelkowski_high-harmonic_2013}, where the interaction between the electron and each atomic center is modeled with a regularized Coulomb potential:
\begin{equation}\label{potmolec}
  V(x)=-\frac{1/2}{\sqrt{\Big(x-\dfrac{R}{2}\Big)^2+a^2}}-\frac{1/2}{\sqrt{\Big(x+\dfrac{R}{2}\Big)^2+a^2}}\,.
\end{equation}
Here $R$ is the internuclear distance, $a$ the regularization parameter and $x$ denotes the electron coordinate along the laser polarization $\vec{e}_x$.
First we will examine the two-center interference condition as a function of $R$, keeping fixed the ionization potential $I_\mathrm{p}$, thus adjusting the regularization parameter $a$ for each value of $R$. We have chosen the one of H$_2$: $I_\mathrm{p}=0.567$ a.u $=15.43$ eV \cite{molecules_Ip}.
Second we will use this model for the specific case of  H$_2$ at equilibrium distance, where $R=1.4$ a.u. with $a=1.1339$ a.u..
The TDSE is solved for an electron in the ground-state interacting with a two-cycle oscillating electric field $\vec{E}(t)=E_\mathrm{L}f(t)\sin(\omega t)\vec{e}_x$, where:
\begin{equation}
  f(t)=
  \left\{
    \begin{array}{ll}
      \sin^2\!\big(\frac{\omega}{4}t\big) & \mbox{if } t \in [0,4\pi/\omega] \\
      0 & \mathrm{elsewhere}.
    \end{array}
  \right.
\end{equation}
Here $E_\mathrm{L}$ denotes the amplitude. Note that for such short pulses, the maximum of the electric field $E_\mathrm{max}\simeq0.87E_\mathrm{L}$ is substantially smaller than the amplitude $E_\mathrm{L}$. To be more consistent with the SFA computations, we define the peak intensity of our pulse as the square of $E_\mathrm{max}$, giving approximately $I_\mathrm{L}\simeq\frac{1}{2}\epsilon_0cE_\mathrm{L}^2/1.322$, with $c$ the speed of light, and $\epsilon_0$ the vacuum permittivity. We have set the frequency $\omega$ to 0.057 a.u., corresponding to a Ti:sapphire laser of 800 nm wavelength.
With such a short IR pulse, the generated spectrum does not exhibit the usual harmonic comb structure: It is continuous through all the plateau region.

Within this model, the TDSE for the electronic wave function $\Psi(x,t)$ reads:
\begin{equation}
  i\frac{\partial}{\partial t}\Psi(x,t)=\bigg[-\frac{1}{2}\frac{\partial^2}{\partial x^2}+V(x)+xE(t)\bigg]\Psi(x,t)\,.
\end{equation}

The HHG spectrum is proportional to the Fourier transform of the time-dependent dipole ${\cal D}(t)$, which is computed as the expectation value of the acceleration at each propagation time-step \cite{burnett_calculation_1992}. While the interest is often put on the harmonic intensity, we focus on its phase as it encodes the dynamic of the system. The dominant contribution to the HHG phase arises from the intrinsic electron dynamics in the continuum (step 2). It is quadratic in energy and evolves rapidly  \cite{mairesse_attosecond_2003,mairesse_optimization_2004}.Thus it might hide the small system-specific signatures  we are interested in. However, the dynamics of the electron freed in the continuum is common to all molecules or atoms of same ionization potential in the same laser conditions. Computations of the HHG spectra fom a reference atom with the {\it same} ionization potential is therefore used to remove this "universal" quadratic phase by subtracting it to the molecular HHG phase \cite{mairesse_attosecond_2003}. Our auxiliary atom is modeled by a single regularized Coulomb potential:
\begin{equation}
  V(x)=-\frac{1}{\sqrt{x^2+b^2}}\,,
\end{equation}
for which we solve the TDSE with the same laser pulse parameters. Here $b=1.2195$ a.u. such that the ionization potential is again $0.567$ a.u.

As the computed spectra contain contributions from both the short and long trajectories, we can use their different spatial properties to separate them and sort out their distinct signatures in the overall HHG phase. Classical computations show that the  excursion  of the electron trajectories in the continuum is characterized by a distance $x_0=E_\mathrm{L}/\omega^2$ \cite{corkum_plasma_1993}. The short trajectories never go beyond $x_0$ while the long trajectories always cross this value. Therefore, we first solved the TDSE with a numerical absorber \cite{krause_calculation_1992} placed at $x_0$, which eliminates the wavefunction components that go further away, corresponding classically to the long trajectories. As a consequence, the surviving part of the wave function is the one of the short trajectories and we can compute the corresponding dipole ${\cal D}_s(t)$. The solution of the same TDSE without the absorber at $x_0$ leads to the full dipole ${\cal D}(t)$. Thence, to recover the long trajectory contribution ${\cal D}_l(t)$, we  subtract ${\cal D}_s(t)$ to ${\cal D}(t)$. One should note that this absorber artificially increases the plateau region leading to a slightly higher cut-off value.

\subsection{Atomic SFA}\label{subsec_methods_SFA}

In the previous section we explained the need of an atomic reference in order to remove the quadratic harmonic phase and collect the molecular specific two-center signature. This stands also in the case of molecular SFA computations, hence, for completeness, we recall in this section the basic equations of atomic SFA \cite{lewenstein_theory_1994}. In SFA computations an abrupt start of the laser field does not affect the spectrum, thus we choose a laser field without envelope:
\begin{equation}\label{eq_laser_SFA}
 \vec{E}(t)=E(t)\vec{e}_x=E_\mathrm{L}\cos(\omega t)\vec{e}_x\,.
\end{equation}
The corresponding vector potential $\vec{A}(t)$ is defined such that:
\begin{equation}
  \vec{E}(t)=-\frac{\partial\vec{A}}{\partial t}\,.
\end{equation}
In the length gauge, the dipole reads:
\begin{multline}\label{eq_dipole_SFA}
  D_\mathrm{at}(\Omega) = \int dt\int_0^t dt'\!\int d^3\vec{p}\,d_{\mathrm{rec}}(\vec{p}+\vec{A}(t)) \\
              \times d_{\mathrm{ion}}(\vec{p}+\vec{A}(t'),t')\e^{-iS(\vec{p},t,t',\Omega)}.
\end{multline}
Here $\Omega$ is the harmonic frequency, $S$ the action,
\begin{equation}
  S(\vec{p},t,t',\Omega)=\int_{t'}^t d\tau\left(\frac{[\vec{p}+\vec{A}(\tau)]^2}{2}+I_\mathrm{p}\right)-\Omega t\,,
\end{equation}
and $d_{\mathrm{ion}}$ and $d_{\mathrm{rec}}$ the projections of respectively the ionization and the recombination dipole matrix elements along the laser polarization, given by: 
      \begin{eqnarray}\label{eq:d_ion_at_def}
        d_{\mathrm{ion}}(\vec{k})&=&\langle\vec{k}|xE(t')|\varphi_0\rangle\nonumber\\
        d_{\mathrm{rec}}(\vec{k})&=&\langle\varphi_0|-i\,\nabla|\vec{k}\rangle\cdot\vec{e}_x,
      \end{eqnarray}
with $|\varphi_0\rangle$ and $|\vec{k}\rangle$ the atomic ground state and continuum wavefunctions, respectively. 

The term $\exp[-iS(\vec{p},t,t',\Omega)]$ is rapidly oscillating. Thus, the five-fold integration in Eq. (\ref{eq_dipole_SFA}) can be reduced to the evaluation of the integrand at the stationary solutions of $S$, for which:
\begin{equation}\label{eq_gradient_action}
 \frac{\partial S}{\partial\vec{p}} = \vec{0},\ \frac{\partial S}{\partial t} = 0,\ \mathrm{and}\ \frac{\partial S}{\partial t'} = 0\,.
\end{equation}
This saddle-point approximation leads to the search of solutions of the following equations:
\begin{subequations}
\label{eq_SP_at}
\begin{alignat}{2}
  & \int_{t'}^t d\tau[\vec{p}+\vec{A}(\tau)] &&= \vec{0}\,, \label{eq_SP_at_p} \\
  & \frac{[\vec{p}+\vec{A}(t)]^2}{2}+I_\mathrm{p}-\Omega &&= 0\,,    \label{eq_SP_at_t} \\
  & \frac{[\vec{p}+\vec{A}(t')]^2}{2}+I_\mathrm{p} &&= 0\,.          \label{eq_SP_at_tp}
\end{alignat}
\end{subequations}
We will denote by $(\vec{p}_\mathrm{at},t_\mathrm{at},t'_\mathrm{at})$ the solutions of Eqs. (\ref{eq_SP_at_p}-\ref{eq_SP_at_tp}) to emphasize that they are related to {\it atomic} trajectories. Note that the stationary momentum can be written:
\begin{equation}\label{pat}
  \vec{p}_\mathrm{at}(t_\mathrm{at},t'_\mathrm{at})=-\frac{\int_{t'_\mathrm{at}}^{t_\mathrm{at}} \vec{A}(\tau)d\tau}{t_\mathrm{at}-t'_\mathrm{at}}\,.
\end{equation}
We also define $S_\mathrm{at}=S(\vec{p}_\mathrm{at},t_\mathrm{at},t'_\mathrm{at},\Omega)$.

The total dipole (Eq. \ref{eq_dipole_SFA}) reduces to:
\begin{multline} \label{eq_dipole_SP_at}
  D_\mathrm{at}(\Omega) = C(t_\mathrm{at},t'_\mathrm{at})d_{\mathrm{rec}}(\vec{p}_\mathrm{at}+\vec{A}(t_\mathrm{at})) \\
              \times d_{\mathrm{ion}}(\vec{p}_\mathrm{at}+\vec{A}(t'_\mathrm{at}),t'_\mathrm{at}) \e^{-iS_\mathrm{at}}\,,
\end{multline}
For the sake of conciseness, no distinction is made between the short or the long trajectories in our notations. As a consequence, the sum over the short and long trajectories in the expression of $D_\mathrm{at}$ will not be \textit{explicitly} indicated in the following. Moreover, and again for clarity, we dropped the dependence of $t_\mathrm{at}$ and $t'_\mathrm{at}$ (and consequently of $\vec{p}_\mathrm{at}$) on $\Omega$. In Eq. (\ref{eq_dipole_SP_at}) $C$ is the saddle-point prefactor which reads \cite{chipperfield_tracking_2006}:
\begin{equation}\label{prefactorat}
  C(t_\mathrm{at},t'_\mathrm{at})=\left(\frac{2\pi}{\epsilon+i(t_\mathrm{at}-t'_\mathrm{at})}\right)^{3/2}\frac{1}{\sqrt{\det(S_p'')|_\mathrm{at}}}\,,
\end{equation}
where $S_p''$ is the Hessian matrix of $S_p(t,t')=S(\vec{p}_\mathrm{at}(t_\mathrm{at},t'_\mathrm{at}),t,t')$:
\begin{equation}
  \det(S_p'')|_\mathrm{at}=\frac{\partial^2 S_p}{\partial t^2}\bigg|_\mathrm{at}\,\frac{\partial^2 S_p}{\partial t'^2}\bigg|_\mathrm{at}-\Bigg(\frac{\partial^2 S_p}{\partial t \partial t'}\bigg|_\mathrm{at}\Bigg)^2\,,
\end{equation}
with $\vec{p}_\mathrm{at}(t_\mathrm{at},t'_\mathrm{at})$ given in Eq. (\ref{pat}). In the prefactor expression given in  Eq. (\ref{prefactorat}), the first factor arises from the evaluation of the integral, Eq. (\ref{eq_dipole_SFA}), over momentum at $\vec{p}_\mathrm{at}$ and accounts for the spreading of the electronic wave-packet in the continuum. The second factor comes from the evaluation of the integral over $t$ and $t'$.

\subsection{Molecular SFA}\label{subsec_methods_SFA_mol}

In this section, we recall the equations of \textit{molecular} SFA as they have been developed in Refs. \cite{chirila_strong-field_2006,faria_high-order_2007}. We will use the expression of the recombination and ionization dipole matrix elements as derived in Ref. \cite{faria_high-order_2007}. These equations are used in the next section (Sec. \ref{subsec_methods_SFA_dl}) to obtain analytical expressions by means of Taylor expansions, which is the key point of our work.

\begin{figure}[h]

  \includegraphics[width=\columnwidth]{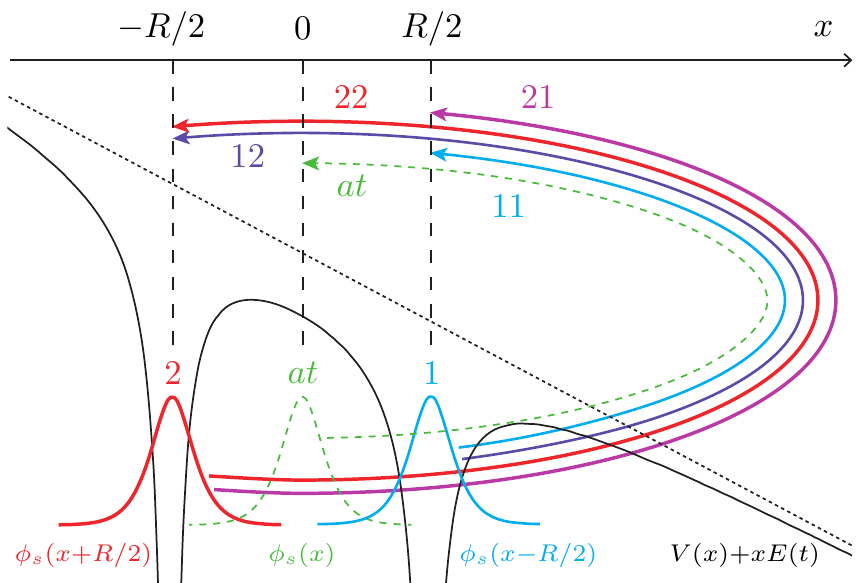}
  \caption{(Color online) One dimensional sketch of  the molecular potential and the four classes of trajectories $\alpha\rightarrow\beta$, with $\alpha,\beta\in\{1,2\}$ in molecular SFA. We present the case of aligned molecules along the laser polarization ($\vec{R}=R\,\vec{e}_x$). We pictured the ionization towards the positive $x$-coordinate, i.e. with a negative laser field $E(t)$. We show in black solid line the resulting potential, sum of $xE(t)$ (dotted line) and of the double-well potential $V(x)$, and the atomic orbitals $\phi_s$ centered on each nucleus used for the computation of the transition dipole matrix elements.}
  \label{fig_traj_SFA_mod}
\end{figure}

The electronic ground state of the target molecule is approximated by a linear combination of atomic orbitals (LCAO), restricted to two atomic orbitals $\phi_s$ centered on the nuclei. Therefore the recombination and ionization dipole matrix elements along $\vec{e}_x$ for a symmetric state ($\sigma_g$) are expressed as \cite{faria_high-order_2007}:
\begin{align}
  & d_{\mathrm{rec}}(\vec{p})=\mathcal{R}(\vec{p})\left(\e^{i\vec{p}\cdot\frac{\vec{R}}{2}}+\e^{-i\vec{p}\cdot\frac{\vec{R}}{2}}\right), \label{eq_d_rec} \\
  & d_{\mathrm{ion}}(\vec{p},t')=\mathcal{I}_1(\vec{p},t')\e^{i\vec{p}\cdot\frac{\vec{R}}{2}}+\mathcal{I}_2(\vec{p},t')\e^{-i\vec{p}\cdot\frac{\vec{R}}{2}}, \label{eq_d_ion}
\end{align}
where:
\begin{align}
  & \mathcal{R}(\vec{p}) = \frac{\bar{\phi}_s(\vec{p})}{\sqrt{2(1+w(R))}}\vec{p}\cdot\vec{e}_x \,, \label{eq_mol_I} \\
  & \mathcal{I}_\alpha(\vec{p},t') = -\frac{\vec{E}(t')}{\sqrt{2\big(1+w(R)\big)}}\cdot\left[i{\pdv{\bar{\phi}_s}{\vec{p}}}(\vec{p})+(-1)^\alpha\frac{\vec{R}}{2}\bar{\phi}_s(\vb{p})\right] \label{eq_mol_R} \,.
\end{align}
Here $\bar{\phi}_s$ is the momentum representation (Fourier transform) of the atomic orbital $\phi_s$ and $w(R)$ the overlap between the two orbitals. In our computations, these orbitals are Gaussian functions \cite{lewenstein_theory_1994}. 
The product of the transition dipole matrix elements thus gives:
\begin{multline}\label{eq_splitting_dion_drec}
  d_{\mathrm{rec}}(\vec{p}+\vec{A}(t))\times d_{\mathrm{ion}}(\vec{p}+\vec{A}(t'),t') = \\
  \sum_{\alpha=1}^2\sum_{\beta=1}^2\mathcal{R}(\vec{p}+\vec{A}(t))\times\mathcal{I}_\alpha(\vec{p}+\vec{A}(t'),t')\e^{-i\Phi_{\alpha\beta}(\vec{p},t,t')}\,,
\end{multline}
where:
\begin{multline}\label{eq_phase_to_action}
  \Phi_{\alpha\beta}(\vec{p},t,t')= \\
  (-1)^\alpha[\vec{p}+\vec{A}(t')]\cdot\frac{\vec{R}}{2}-(-1)^\beta[\vec{p}+\vec{A}(t)]\cdot\frac{\vec{R}}{2}\,.
\end{multline}
Here $\alpha$ and $\beta$ label the initial and final atomic center in the electron excursion, respectively.
Thus, the total \textit{molecular} dipole reads:
\begin{multline}\label{eq_dipole_sfa_mol_no_sdp}
  D(\Omega) =\sum_{\alpha=1}^2\sum_{\beta=1}^2\int dt\int_0^t dt'\!\int d\vec{p}\,\mathcal{R}(\vec{p}+\vec{A}(t)) \\
              \times\mathcal{I}_\alpha(\vec{p}+\vec{A}(t'),t')\e^{-iS_{\alpha\beta}(\vec{p},t,t',\Omega)}\,.
\end{multline}
It is the sum of four terms that can be evaluated independently by searching the stationary points of the \textit{modified} actions:
\begin{equation}\label{eq_mod_action}
  S_{\alpha\beta}(\vec{p},t,t',\Omega)=S(\vec{p},t,t',\Omega)+\Phi_{\alpha\beta}(\vec{p},t,t')\,.
\end{equation}
This leads to the following saddle-point equations \cite{faria_high-order_2007}:
\begin{subequations}
\label{eq_SP_mol}
\begin{alignat}{2}
   & \displaystyle\int_{t'}^t d\tau[\vec{p}+\vec{A}(\tau)] + (-1)^\alpha\frac{\vec{R}}{2}-(-1)^\beta\frac{\vec{R}}{2} &&= \vec{0}\,, \label{eq_SP_mol_p} \\
  & \frac{[\vec{p}+\vec{A}(t)]^2}{2}+I_\mathrm{p}-\Omega+(-1)^\beta \vec{E}(t)\cdot\frac{\vec{R}}{2} &&= 0\,, \label{eq_SP_mol_t} \\
  & \frac{[\vec{p}+\vec{A}(t')]^2}{2}+I_\mathrm{p}+(-1)^\alpha \vec{E}(t')\cdot\frac{\vec{R}}{2} &&= 0\,, \label{eq_SP_mol_tp}
\end{alignat}
\end{subequations}
for which we find four new groups of stationary solutions $(\vec{p}_{\alpha\beta},t_{\alpha\beta},t'_{\alpha\beta})$. They correspond to four classes of trajectories, represented in Fig. \ref{fig_traj_SFA_mod}. They express the ionization from one center and the recombination to either the same center or the other one. Each class contains a short and a long path. Within the saddle-point approximation, the total dipole thus reduces to:
\begin{multline}\label{eq_tot_dip_mod_sdp}
  D(\Omega) =\sum_{\alpha=1}^2\sum_{\beta=1}^2C_{\alpha\beta}(t_{\alpha\beta},t'_{\alpha\beta})\mathcal{R}(\vec{p}_{\alpha\beta}+\vec{A}(t_{\alpha\beta})) \\
              \times \mathcal{I}_\alpha(\vec{p}_{\alpha\beta}+\vec{A}(t'_{\alpha\beta}),t'_{\alpha\beta})\e^{-iS_{\alpha\beta}(\vec{p}_{\alpha\beta},t_{\alpha\beta},t'_{\alpha\beta},\Omega)},
\end{multline}
with $C_{\alpha\beta}(t_{\alpha\beta},t'_{\alpha\beta})$ the resulting saddle-point prefactors, see Eq.(\ref{prefactorat}).

We will show in Sec. \ref{sec_results} that the observations on the phase-jumps in the ab initio computations are well reproduced within the molecular SFA. This motivated us to further analyze the molecular SFA analytically and get a proper understanding of these behaviors. In the following, we express the deviations of the molecular solutions regarding to the reference atomic ones by means of Taylor expansions.

\subsection{Taylor expansion of the molecular SFA}\label{subsec_methods_SFA_dl}

\subsubsection{Stationary points}

\begin{table}
  \begin{tabular}{ccllll}
    \hline
      & & \multicolumn{2}{c}{Short trajectories} & \multicolumn{2}{c}{Long trajectories} \\
    \hline
      $ $ & H & \multicolumn{1}{c}{Re} & \multicolumn{1}{c}{Im} & \multicolumn{1}{c}{Re} & \multicolumn{1}{c}{Im} \\
    \hline
    \hline
    $p_{11}$    & 21 & 1.8976  & 0.14767  & 0.091138 &{ }5.0125$\times10^{-4}$ \\
                & 31 & 1.6986  & 0.089790 & 0.14357  &{ }7.7589$\times10^{-4}$ \\
                & 41 & 1.5214  & 0.061121 & 0.20458  &{ }1.0670$\times10^{-3}$ \\
                & 51 & 1.3508  & 0.044192 & 0.27847  &{ }1.3447$\times10^{-3}$ \\
                & 61 & 1.1706  & 0.033242 & 0.37518  &{ }1.4493$\times10^{-3}$ \\
                & 71 & 0.93811 & 0.027673 & 0.53362  &  -5.5265$\times10^{-4}$ \\
    \hline
    $p_{12}$    & 21 & 1.8976  & 0.14769  & 0.091138 &{ }5.0125$\times10^{-4}$ \\
                & 31 & 1.6986  & 0.089793 & 0.14357  &{ }7.7590$\times10^{-4}$ \\
                & 41 & 1.5214  & 0.061122 & 0.20458  &{ }1.0670$\times10^{-3}$ \\
                & 51 & 1.3508  & 0.044192 & 0.27847  &{ }1.3447$\times10^{-3}$ \\
                & 61 & 1.1706  & 0.033242 & 0.37518  &{ }1.4493$\times10^{-3}$ \\
                & 71 & 0.93811 & 0.027673 & 0.53362  &  -5.5264$\times10^{-4}$ \\
    \hline
    $p_{21}$    & 21 & 1.8975  & 0.14772  & 0.091138 &{ }5.0176$\times10^{-4}$ \\
                & 31 & 1.6986  & 0.089830 & 0.14357  &{ }7.7668$\times10^{-4}$ \\
                & 41 & 1.5214  & 0.061154 & 0.20458  &{ }1.0681$\times10^{-3}$ \\
                & 51 & 1.3508  & 0.044219 & 0.27847  &{ }1.3460$\times10^{-3}$ \\
                & 61 & 1.1706  & 0.033266 & 0.37518  &{ }1.4507$\times10^{-3}$ \\
                & 71 & 0.93811 & 0.027695 & 0.53362  &  -5.5317$\times10^{-4}$ \\
    \hline
    $p_{22}$    & 21 & 1.8976  & 0.14774  & 0.091138 &{ }5.0176$\times10^{-4}$ \\
                & 31 & 1.6986  & 0.089833 & 0.14357  &{ }7.7668$\times10^{-4}$ \\
                & 41 & 1.5214  & 0.061155 & 0.20458  &{ }1.0681$\times10^{-3}$ \\
                & 51 & 1.3508  & 0.044220 & 0.27847  &{ }1.3460$\times10^{-3}$ \\
                & 61 & 1.1706  & 0.033266 & 0.37518  &{ }1.4507$\times10^{-3}$ \\
                & 71 & 0.93811 & 0.027695 & 0.53362  &  -5.5316$\times10^{-4}$ \\
    \hline
    $p_{2^\mathrm{nd}}$   & 21 & 1.8976  & 0.14770  & 0.091138  &{ }5.0152$\times10^{-4}$ \\
                          & 31 & 1.6986  & 0.089812 & 0.14357   &{ }7.7631$\times10^{-4}$ \\
                          & 41 & 1.5214  & 0.061138 & 0.20458   &{ }1.0676$\times10^{-3}$ \\
                          & 51 & 1.3508  & 0.044206 & 0.27847   &{ }1.3454$\times10^{-3}$ \\
                          & 61 & 1.1706  & 0.033255 & 0.37518   &{ }1.4501$\times10^{-3}$ \\
                          & 71 & 0.93811 & 0.027685 & 0.53362   &  -5.5280$\times10^{-4}$ \\
    \hline
    $p_\mathrm{at}$     & 21 & 1.8972  & 0.14788  & 0.091148  &{ }5.0516$\times10^{-4}$ \\
    $=p_{1^\mathrm{st}}$& 31 & 1.6982  & 0.089978 & 0.14359   &{ }7.8195$\times10^{-4}$ \\
                        & 41 & 1.5210  & 0.061302 & 0.20461   &{ }1.0753$\times10^{-3}$ \\
                        & 51 & 1.3504  & 0.044360 & 0.27853   &{ }1.3549$\times10^{-3}$ \\
                        & 61 & 1.1702  & 0.033397 & 0.37528   &{ }1.4596$\times10^{-3}$ \\
                        & 71 & 0.93754 & 0.027841 & 0.53390   &  -5.6899$\times10^{-4}$ \\
  \end{tabular}
  \caption{ \label{table_p_deviations} Numerical values of the stationary momenta $p_{\alpha\beta}$ solution of the saddle point Eqs. (\ref{eq_SP_mol}), compared to its development Eq. (\ref{eq_sol_mol_DL}), restricted either to $1^{\mathrm{st}}$ or $2^{\mathrm{nd}}$ order, for six harmonics (column denoted by the H symbol) within the plateau of the spectrum generated at a laser intensity of $5\times10^{14}$ W.cm$^{-2}$.}
\end{table}

\begin{table*}[t]
  \begin{tabular}{ccccccc@{\hskip 11pt}ccccc}
  \hline
    & & \multicolumn{5}{c@{\hskip 11pt}}{Re($t_{\alpha\beta}-t_\mathrm{at}$)} & \multicolumn{5}{c}{Im($t'_{\alpha\beta}-t'_\mathrm{at}$)} \\
  \hline
    & & \multicolumn{2}{c}{Numerical} & \multicolumn{2}{c}{2nd order} & \multicolumn{1}{c@{\hskip 11pt}}{1st order} & \multicolumn{2}{c}{Numerical} & \multicolumn{2}{c}{2nd order} & \multicolumn{1}{c}{1st order} \\
  \hline
    $\alpha\beta$ & H & S & L & S & L & \multicolumn{1}{c@{\hskip 11pt}}{} & S & L & S & L \\
  \hline
   11 & 21 &   -0.643 &   -0.601 &   -0.639 &   -0.599 &   -0.624 &   -0.669 &   -0.681 &   -0.668 &   -0.679 &   -0.657 \\
      & 31 &   -0.463 &   -0.441 &   -0.463 &   -0.441 &   -0.452 &   -0.670 &   -0.681 &   -0.670 &   -0.679 &   -0.657 \\
      & 41 &   -0.381 &   -0.365 &   -0.381 &   -0.365 &   -0.372 &   -0.672 &   -0.681 &   -0.672 &   -0.679 &   -0.657 \\
      & 51 &   -0.331 &   -0.317 &   -0.331 &   -0.317 &   -0.324 &   -0.674 &   -0.681 &   -0.673 &   -0.679 &   -0.657 \\
      & 61 &   -0.298 &   -0.284 &   -0.298 &   -0.284 &   -0.290 &   -0.676 &   -0.681 &   -0.675 &   -0.679 &   -0.657 \\
      & 71 &   -0.277 &   -0.255 &   -0.277 &   -0.255 &   -0.265 &   -0.678 &   -0.680 &   -0.677 &   -0.680 &   -0.657 \\
  \hline
   12 & 21 & { }0.605 & { }0.650 & { }0.608 & { }0.648 & { }0.624 &   -0.668 &   -0.681 &   -0.668 &   -0.679 &   -0.657 \\
      & 31 & { }0.441 & { }0.463 & { }0.441 & { }0.463 & { }0.452 &   -0.670 &   -0.681 &   -0.670 &   -0.679 &   -0.657 \\
      & 41 & { }0.364 & { }0.380 & { }0.363 & { }0.379 & { }0.372 &   -0.672 &   -0.681 &   -0.672 &   -0.679 &   -0.657 \\
      & 51 & { }0.316 & { }0.330 & { }0.316 & { }0.330 & { }0.324 &   -0.674 &   -0.681 &   -0.673 &   -0.679 &   -0.657 \\
      & 61 & { }0.282 & { }0.297 & { }0.282 & { }0.297 & { }0.290 &   -0.676 &   -0.681 &   -0.675 &   -0.679 &   -0.657 \\
      & 71 & { }0.254 & { }0.276 & { }0.253 & { }0.276 & { }0.265 &   -0.678 &   -0.680 &   -0.677 &   -0.680 &   -0.657 \\
  \hline
   21 & 21 &   -0.642 &   -0.601 &   -0.639 &   -0.599 &   -0.624 & { }0.647 & { }0.637 & { }0.646 & { }0.636 & { }0.657 \\
      & 31 &   -0.463 &   -0.441 &   -0.463 &   -0.441 &   -0.452 & { }0.646 & { }0.637 & { }0.645 & { }0.636 & { }0.657 \\
      & 41 &   -0.381 &   -0.365 &   -0.381 &   -0.365 &   -0.372 & { }0.644 & { }0.637 & { }0.643 & { }0.636 & { }0.657 \\
      & 51 &   -0.331 &   -0.317 &   -0.331 &   -0.317 &   -0.324 & { }0.642 & { }0.637 & { }0.641 & { }0.636 & { }0.657 \\
      & 61 &   -0.298 &   -0.284 &   -0.298 &   -0.284 &   -0.290 & { }0.641 & { }0.637 & { }0.640 & { }0.636 & { }0.657 \\
      & 71 &   -0.277 &   -0.255 &   -0.277 &   -0.255 &   -0.265 & { }0.639 & { }0.638 & { }0.638 & { }0.636 & { }0.657 \\
  \hline
   22 & 21 & { }0.605 & { }0.650 & { }0.608 & { }0.648 & { }0.624 & { }0.647 & { }0.637 & { }0.646 & { }0.636 & { }0.657 \\
      & 31 & { }0.441 & { }0.463 & { }0.441 & { }0.463 & { }0.452 & { }0.646 & { }0.637 & { }0.645 & { }0.636 & { }0.657 \\
      & 41 & { }0.364 & { }0.380 & { }0.363 & { }0.379 & { }0.372 & { }0.644 & { }0.637 & { }0.643 & { }0.636 & { }0.657 \\
      & 51 & { }0.316 & { }0.330 & { }0.316 & { }0.330 & { }0.324 & { }0.642 & { }0.637 & { }0.641 & { }0.636 & { }0.657 \\
      & 61 & { }0.283 & { }0.297 & { }0.282 & { }0.297 & { }0.290 & { }0.641 & { }0.637 & { }0.640 & { }0.636 & { }0.657 \\
      & 71 & { }0.254 & { }0.276 & { }0.253 & { }0.276 & { }0.265 & { }0.639 & { }0.638 & { }0.638 & { }0.636 & { }0.657 \\
  \hline
\end{tabular}
  \caption{ \label{table_time_deviations} Numerical values of the time deviations $t_{\alpha\beta}-t_\mathrm{at}$ and $t'_{\alpha\beta}-t_\mathrm{at}$ for the same harmonics and same generation conditions as in Table \ref{table_p_deviations}, compared with our development \eqref{eq_sol_mol_DL}, to either the $1^{\mathrm{st}}$ or the $2^{\mathrm{nd}}$ order. We present the numerical values for the short (S) and the long (L) trajectories. We do not present the imaginary part of $t_{\alpha\beta}-t_\mathrm{at}$ ($<3\times10^{-3}$ a.u.), nor the real part of $t'_{\alpha\beta}-t'_\mathrm{at}$ ($<5\times10^{-3}$ a.u.). The  deviations from the atomic case are small compared to the imaginary part of the ionization times ($7.83-15.8$ a.u.) and the real part of the recollision times ($32.9-107$ a.u.), respectively.}
\end{table*}

As we explained, in Sec. \ref{subsec_methods_ab_initio} we will calibrate our numerical results with the atomic ones in order to remove the quadratic harmonic phase. Moreover, since internuclear distances (few \AA) are small compared to the travel path of the freed electrons in the continuum  (several tens of \AA), the times for molecular trajectories are expected to remain close to the ones of an atomic trajectory. These considerations conducted us to perform Taylor expansions in powers of the internuclear distance $R$ of the molecular SFA equations, Eq. (\ref{eq_SP_mol}). To remain consistent with our TDSE simulations, where we restricted the study to a 1D case, we will consider in the following, the specific case of a model molecule, with only one degree of freedom, aligned along the laser polarization,  i.e. $\vec{R}=R\,\vec{e}_x$ and $\vec{p}=p\,\vec{e}_x$.  Moreover, the "natural" absence of a perpendicular component of the momentum allows us to derive equations which reveal straightforward physical insights.

The solutions of the molecular saddle point equations (\ref{eq_SP_mol}) are expanded as:
\begin{subequations}
\label{eq_sol_mol_DL}
\begin{align}
p_{\alpha\beta}  & = p_\mathrm{at}  + \Delta p^{(1)}_{\alpha\beta} + \Delta p^{(2)}_{\alpha\beta} + O(R^3), \\
t_{\alpha\beta}  & = t_\mathrm{at}  + \Delta t^{(1)}_{\alpha\beta} + \Delta t^{(2)}_{\alpha\beta} + O(R^3), \\
t'_{\alpha\beta} & = t'_\mathrm{at} + \Delta t'^{(1)}_{\alpha\beta}+ \Delta t'^{(2)}_{\alpha\beta}+ O(R^3), 
\end{align}
\end{subequations}
where $\Delta p^{(1)}_{\alpha\beta}$, $\Delta t^{(1)}_{\alpha\beta}$ and $\Delta t'^{(1)}_{\alpha\beta}$ are proportiona to $R$ and $\Delta p^{(2)}_{\alpha\beta}$, $\Delta t^{(2)}_{\alpha\beta}$ and $\Delta t'^{(2)}_{\alpha\beta}$ are proportional to $R^2$. Inserting this in Eq.~(\ref{eq_SP_mol}), we obtain a linear set of equations for the first order terms :
\begin{subequations}
\label{eq_SP_mol_1st}
\begin{align}
\label{eq_SP_mol_1st_p}
0= &\left[p_\mathrm{at}+ A(t_\mathrm{at})\right]\Delta t^{(1)}_{\alpha\beta} - \left[p_\mathrm{at}+ A(t'_\mathrm{at})\right]\Delta t'^{(1)}_{\alpha\beta}\nonumber \\ 
   &+ (t_\mathrm{at}-t'_\mathrm{at})\Delta p^{(1)}_{\alpha\beta} + \frac{(-1)^\alpha-(-1)^\beta}{2}R,  \\
\label{eq_SP_mol_1st_t}
0= &-\left[p_\mathrm{at}+ A(t_\mathrm{at})\right]E(t_\mathrm{at})\Delta t^{(1)}_{\alpha\beta} + \left[p_\mathrm{at}+ A(t_\mathrm{at})\right]\Delta p^{(1)}_{\alpha\beta} \nonumber \\
   &+ \frac{(-1)^\beta}{2}E(t_\mathrm{at})R,  \\
\label{eq_SP_mol_1st_tp}
0= &-\left[p_\mathrm{at}+ A(t'_\mathrm{at})\right]E(t'_\mathrm{at})\Delta t'^{(1)}_{\alpha\beta} + \left[p_\mathrm{at}+ A(t'_\mathrm{at})\right]\Delta p^{(1)}_{\alpha\beta} \nonumber \\
  &+ \frac{(-1)^\alpha}{2}E(t'_\mathrm{at})R.   
\end{align}
\end{subequations}
Resolving this set of equations (\ref{eq_SP_mol_1st}) gives:
\begin{subequations}
\label{eq_SP_mol_1st_sol}
\begin{align}
\label{eq_SP_mol_1st_sol_p}
\Delta p^{(1)}_{\alpha\beta}&=0, \\
\label{eq_SP_mol_1st_sol_t}
\Delta t^{(1)}_{\alpha\beta}&=\frac{(-1)^\beta R}{2\left[p_\mathrm{at}+ A(t_\mathrm{at})\right]}, \\
\label{eq_SP_mol_1st_sol_tp}
\Delta t'^{(1)}_{\alpha\beta}&=\frac{(-1)^\alpha R}{2\left[p_\mathrm{at}+ A(t'_\mathrm{at})\right]}.
\end{align}
\end{subequations}
Using, Eqs. (\ref{eq_SP_at}), the last two equations can be rewritten as \footnote{In these  equations and in their interpretation below, we used the sign convention illustrated in Fig. \eqref{fig_traj_SFA_mod}, where the barrier is lowered for $x >0$, i.e. when the electric field is negative at ionization time.}
\begin{subequations}
\label{eq_SP_mol_1st_solb}
\begin{align}
\label{eq_SP_mol_1st_sol_tb}
\Delta t^{(1)}_{\alpha\beta}&=\frac{(-1)^\beta R/2}{\sqrt{2(\Omega-I_p)}}, \\
\label{eq_SP_mol_1st_sol_tpb}
\Delta t'^{(1)}_{\alpha\beta}&=i\frac{(-1)^\alpha R/2}{\sqrt{2I_p}}.
\end{align}
\end{subequations}
The same procedure can be performed for the second order terms, see Appendix A.

Table \ref{table_p_deviations} and \ref{table_time_deviations} display the numerical values of the stationary momentum $p_{\alpha\beta}$ and of the ionization and recollision time deviations $t_{\alpha\beta}-t_\mathrm{at}$ and $t'_{\alpha\beta}-t'_\mathrm{at}$ and compare them with the values from the developments including either the first order term {\it only}  or both the first and the second order. An excellent agreement is found between the numerically exact results and the development including up to the second order. Yet, we note that the first order term is in very good agreement with the exact values, with errors limited to a few percent (around 0.02 a.u.). We will therefore restrict our development to this order. As will be shown, the first order treatment has the valuable advantage to provide
a direct and simple physical picture on the underlying timings of ionization and recollision in diatomic molecules.

First, the ionization time is shifted by a {\it pure imaginary} delay $\Delta t'^{(1)}_{\alpha\beta}$, as seen in Eq. (\ref{eq_SP_mol_1st_sol_tpb}), which can be interpreted as a correction to the tunneling time with respect to the atomic case. If the electron is ionized from the center $\alpha=1$, located at $R/2$, we find that $\mathrm{Im}(\Delta t'^{(1)}_{1\beta})<0$ which means that the time spend “below the barrier” is smaller. This is consistent with a smaller barrier to cross, as seen in Fig. \ref{fig_traj_SFA_mod}. The opposite situation is observed for an electron ionized from center $\alpha=2$ for which $\mathrm{Im}(\Delta t'^{(1)}_{2\beta})>0$, where a larger barrier is to be  crossed. 

Second, the shift for the recollision times with respect to the atomic times, $\Delta t^{(1)}_{\alpha\beta}$, Eq. (\ref{eq_SP_mol_1st_sol_tb}), are now real. For an electron recombining with the center located at $R/2$ again, i.e. $\beta=1$, we find that $\Delta t^{(1)}_{\alpha 1}<0$ which means that the electron recollides earlier compared to the atomic case. Indeed, as pictured in Fig. \ref{fig_traj_SFA_mod}, the electron travels a shorter path. On the contrary, an electron recolliding with center $\beta=2$ for which $\Delta t^{(1)}_{\alpha 2}>0$ travels a longer path. 

Note that for both time delays, $\Delta t^{(1)}_{\alpha\beta}$ and $\Delta t'^{(1)}_{\alpha\beta}$, their formal expressions, Eqs. (\ref{eq_SP_mol_1st_sol}),  can be regarded as the ratio between the additional distance to cross in molecule, $\pm R/2$, and the velocity of the electron given in Eqs. \ref{eq_SP_at_t}-\ref{eq_SP_at_tp}, either classically allowed or forbidden.

We now turn to the Taylor expansion of the different terms entering the expression of the total dipole in Eq. (\ref{eq_tot_dip_mod_sdp}) in order to find out how these differences of timing affect the phase of HHG.

\subsubsection{Total dipole}\label{subsubsec_total}

We first expand the modified action $S_{\alpha\beta}$ to first-order in $R$:
\begin{multline}\label{eq_S_mol_dl}
  S_{\alpha\beta}(p_{\alpha\beta},t_{\alpha\beta},t'_{\alpha\beta},\Omega) = S_\mathrm{at}+ \Phi_{\alpha\beta}(p_\mathrm{at},t_\mathrm{at},t'_\mathrm{at})\\+\left.\frac{\partial S}{\partial t}\right|_\mathrm{at} \Delta t^{(1)}_{\alpha\beta}+\left.\frac{\partial S}{\partial t'}\right|_\mathrm{at}\Delta t'^{(1)}_{\alpha\beta}  + O(R^2).
\end{multline}
The partial derivatives of the action $S$ with respect to $t$ and $t'$ cancel at the atomic stationary points (\ref{eq_gradient_action}), so we simply get:
\begin{multline}
  S_{\alpha\beta}(p_{\alpha\beta},t_{\alpha\beta},t'_{\alpha\beta},\Omega) =  S_\mathrm{at}+ \Phi_{\alpha\beta}(p_\mathrm{at},t_\mathrm{at},t'_\mathrm{at}) + O(R^2).
\end{multline}
Then we expand the ionization and recombination dipole matrix elements (Eqs. \ref{eq_mol_I} and \ref{eq_mol_R}) :
\begin{multline}\label{eq_Rab}
  \mathcal{R}(p_{\alpha\beta}+A(t_{\alpha\beta}))=\mathcal{R}_\mathrm{at}(p_\mathrm{at}+A(t_\mathrm{at}))\\
  +(-1)^\beta \mathcal{R}^{(1)}(p_\mathrm{at}+A(t_\mathrm{at}),t_\mathrm{at})\,R +O(R^2) \,,
\end{multline}
where:
\begin{align}\label{eq_R1}
  \mathcal{R}_\mathrm{at}(p)&=\frac{p\bar{\phi}_s(p)}{2},\\
  \mathcal{R}^{(1)}(p,t)&=-\frac{E(t)}{4}\left(\frac{\bar{\phi}_s(p)}{p}+\frac{\partial\bar{\phi}_s}{\partial p}(p)\right)
\end{align}
and
\begin{multline}\label{eq_Iab}
  \mathcal{I}_\alpha(p_{\alpha\beta}+A(t'_{\alpha\beta}),t'_{\alpha\beta})=\mathcal{I}_\mathrm{at}(p_\mathrm{at}+A(t'_\mathrm{at}),t'_\mathrm{at}) \\- (-1)^\alpha \mathcal{I}^{(1)}(p_\mathrm{at}+A(t'_\mathrm{at}),t'_\mathrm{at}) \,R +O(R^2)\,,
\end{multline}
where:
\begin{align}\label{eq_I1}
  \mathcal{I}_\mathrm{at}(p,t')=&\frac{-iE(t')}{2}\frac{\partial\bar{\phi}_s}{\partial p}(p), \\
  \mathcal{I}^{(1)}(p,t')=&\frac{E(t')}{4}\bar{\phi}_s(p)+\frac{i\omega^2A(t')}{4p}\frac{\partial \bar{\phi}_s}{\partial p}(p) \nonumber\\&-i\frac{E^2(t')}{4p}\frac{\partial^2 \bar{\phi}_s}{\partial p^2}(p).
\end{align}

We checked numerically that the molecular prefactor $C_{\alpha\beta}(t_{\alpha\beta},t'_{\alpha\beta})$ is actually very close to the atomic one. We will thus, as a first approximation, consider it to be equal to $C(t_\mathrm{at},t'_\mathrm{at})$ (\ref{prefactorat}). The exact effect of this approximation will be discussed in Sec. \ref{subsec_interp_extra_phase}. 

Finally, after regrouping all the terms, the first order approximation of the total dipole (\ref{eq_tot_dip_mod_sdp}) can be refolded into:
\begin{align}\label{eq_dipole_mol_dl}
\widetilde{D}(\Omega)=&C(t_\mathrm{at},t'_\mathrm{at})\widetilde{d}_{\mathrm{rec}}(p_\mathrm{at}+A(t_\mathrm{at}),t_\mathrm{at}) \nonumber \\
& \times \widetilde{d}_{\mathrm{ion}}(p_\mathrm{at}+A(t'_\mathrm{at}),t'_\mathrm{at}) e^{-iS_\mathrm{at}}\,.
\end{align}
We have thus retrieved an expression which is similar to the atomic dipole given by Eq. (\ref{eq_dipole_SFA}). It is evaluated at atomic stationary points $(p_\mathrm{at},t_\mathrm{at},t'_\mathrm{at})$ and involves the atomic action $S_\mathrm{at}$. However, it involves \textit{modified} transition dipole matrix elements $\widetilde{d}_{\mathrm{ion}}$ and $\widetilde{d}_{\mathrm{rec}}$ expressed as:
\begin{multline}\label{eq_drec_tilde}
  \widetilde{d}_{\mathrm{rec}}(p,t)=2\mathcal{R}(p)\cos\left(\frac{pR}{2}\right) \\
  +2i\,\mathcal{R}^{(1)}(p,t)R\sin\left(\frac{pR}{2}\right),
\end{multline}
\begin{multline}\label{eq_dion_tilde}
  \widetilde{d}_{\mathrm{ion}}(p,t')=2\mathcal{I}_\mathrm{at}(p,t')\cos\left(\frac{pR}{2}\right) \\
  +2i\,\mathcal{I}^{(1)}(p,t')R\sin\left(\frac{pR}{2}\right).
\end{multline}
These expressions are reminiscent to the ones of \cite{spiewanowski_high-order-harmonic_2013,spiewanowski_field-induced_2014}, where field dressed states were used a priori to define the dipole.
\section{Numerical simulations}\label{sec_results}

\subsection{Ab initio computations}\label{subsec_results_ab_initio}

\begin{figure}
\begin{center}
 \includegraphics[width=\columnwidth]{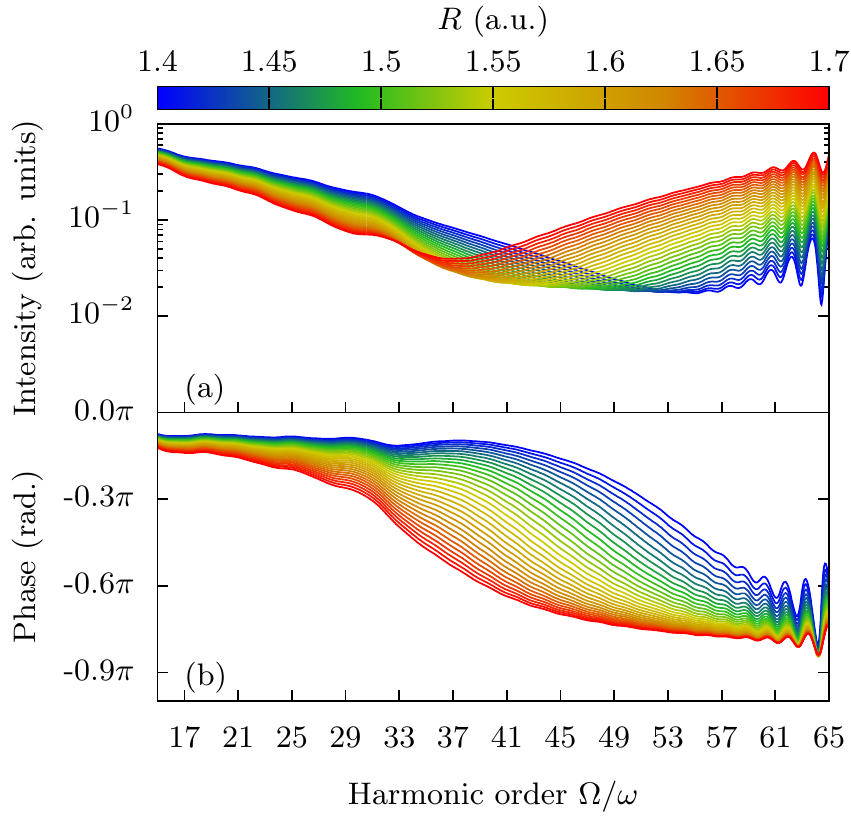}
  \caption{(Color online) Ab initio computations of the contribution of the short trajectories to high-order harmonic spectra for a two-cycle laser pulse of $2.65\times10^{14}$ W.cm$^{-2}$ peak intensity for a diatomic molecule with different internuclear distances $R$ from 1.4 to 1.7 a.u. We report the harmonic intensity (a) and phase (b) calibrated to an atomic reference.}
  \label{fig_TDSE_ampl_phase_short_R}
\end{center}
\end{figure}

\begin{figure}
\begin{center}
 \includegraphics[width=\columnwidth]{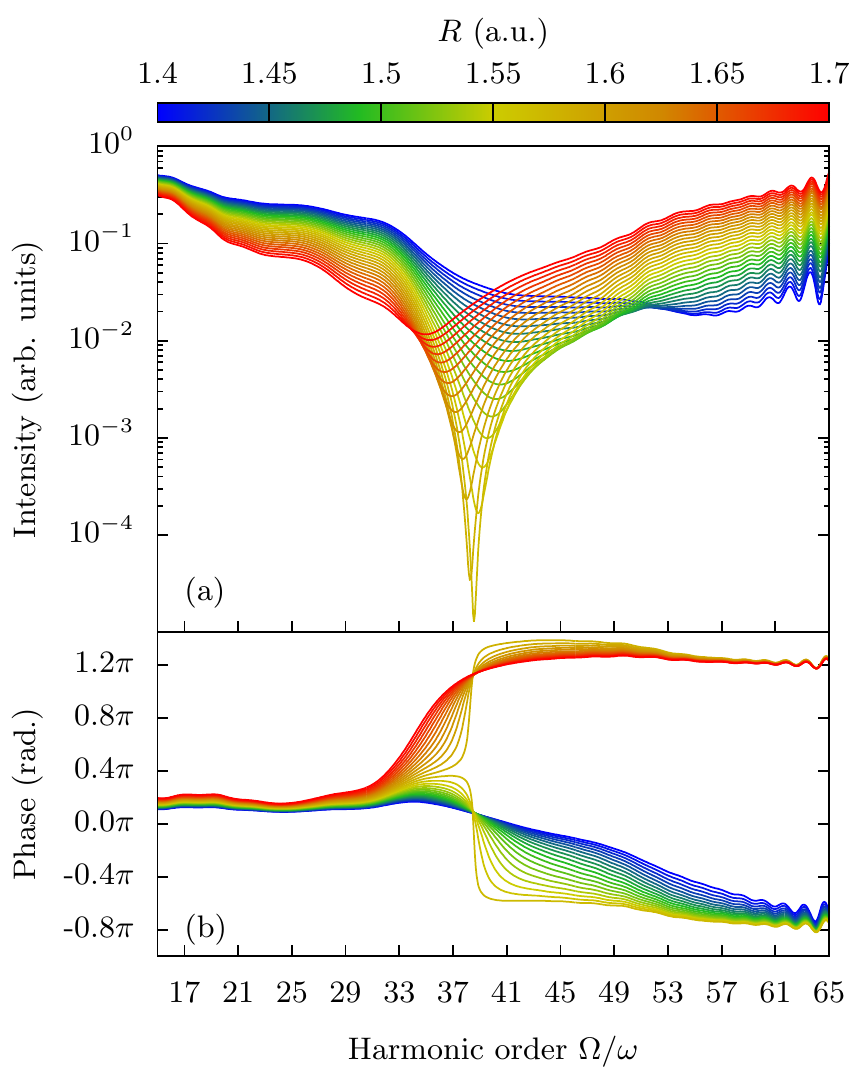}
  \caption{(Color online) Same as in Fig. \ref{fig_TDSE_ampl_phase_short_R} for the long trajectories.}
  \label{fig_TDSE_ampl_phase_long_R}
\end{center}
\end{figure}

As presented in Sec. \ref{subsec_methods_ab_initio} we first investigate the signature of two-center interference in both the short and long trajectory contributions to the HHG spectrum with our TDSE computations. The results are presented in Figs. \ref{fig_TDSE_ampl_phase_short_R} and \ref{fig_TDSE_ampl_phase_long_R}. We report the intensity and phase of the short and long trajectory contributions to HHG spectra for several internuclear distances between 1.4 and 1.7 a.u., generated at a laser peak intensity of $2.65\times10^{14}$ W.cm$^{-2}$. The calibration with a reference atom, as explained in Sec. \ref{sec_methods}, implies to divide the harmonic intensity by the one of the atom. Consistently, the reported phase is the phase difference between the molecular and the atomic computations, which removes the quadratic contribution and allows the observation of the two-center interference effects.

In the intensity of each spectrum we observe a minimum, which is shallow in the case of the short trajectories (Fig. \ref{fig_TDSE_ampl_phase_short_R}a), and much more pronounced for the long trajectories, especially for particular values of $R$ (Fig. \ref{fig_TDSE_ampl_phase_long_R}a). This minimum is the signature of the destructive interference induced by the two centers of the molecule  we are interested in. Its position has long been ascribed to the zeros of the recombination dipole matrix element $d_{\mathrm{rec}}$ given by Eq. (\ref{eq_d_rec}), which relies on plane-waves (PW) to  describe the continuum and on the LCAO approximation for the ground state \cite{lein_interference_2002}. Within these approximations, the zeros of $d_{\mathrm{rec}}$ occur at particular momentum values $p_k=(2k+1)\pi/R$, $k\in\mathbb{Z}$. We focus on the first value $p_0$ as it is the only one contained in our harmonic spectra generated by a relatively short-wavelength laser like Ti:sapphire. The corresponding harmonic frequency is:
\begin{equation}\label{eq_omega0}
  \Omega_0^{(\mathrm{PW})} = I_\mathrm{p}+\frac{p_0^2}{2} = I_\mathrm{p}+\frac{\pi^2}{2R^2}\,.
\end{equation}
For the illustrative case $R=1.57$ a.u., this lead to a minimum at harmonic 45, while our TDSE simulations give a minimum at "harmonic" 38.5.
 
A way to overcome this  discrepancy is to claim that effective PW representing the returning electron must be energy-shifted by $I_\mathrm{p}$ \cite{lein_interference_2002}, leading to the approximate expression:
\begin{equation}\label{eq_omega0_TDSE}
  \bar{\Omega}_0\simeq \frac{\pi^2}{2R^2}\,.
\end{equation}
Still, this correction is not sufficient as it predicts a minimum at harmonic 35. 

\begin{figure}
\begin{center}
 \includegraphics[width=\columnwidth]{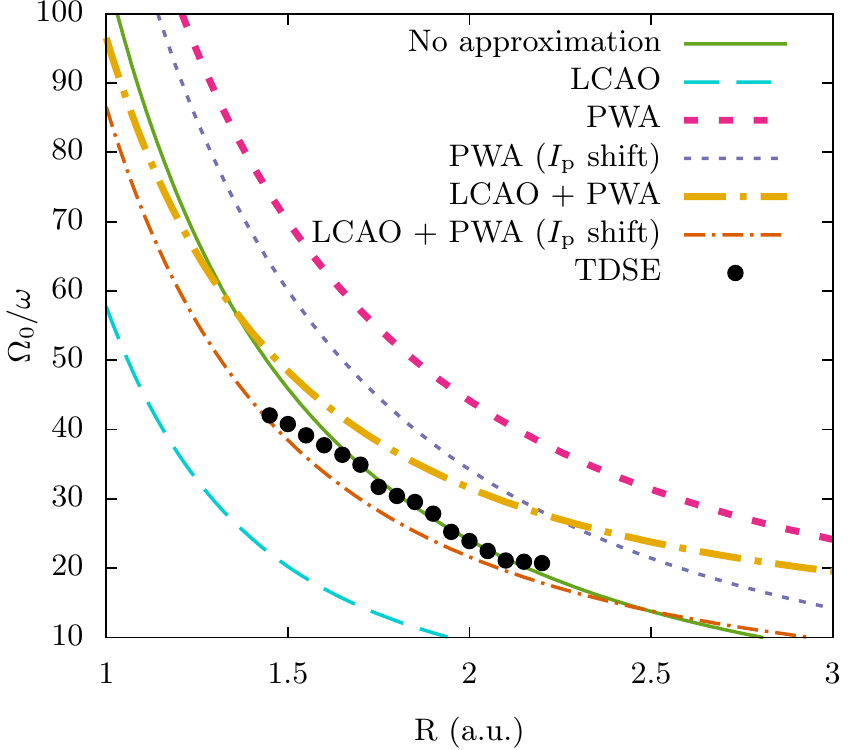}
  \caption{(Color online) Harmonic order of the minimum as a function of $R$. Black dots: minimum in the harmonic spectrum extracted from TDSE calculations, lines: zero of the recombination dipole matrix element computed at different levels of approximation.}
  \label{fig_minimum_position}
\end{center}
\end{figure}

Therefore, we used numerical simulations to better understand why the analytical formula, either with or without the $I_\mathrm{p}$ shift, respectively Eq. (\ref{eq_omega0_TDSE}) and (\ref{eq_omega0}), does not reproduce the position of the minimum. We computed numerically the eigenstates of the field-free system by solving the time independent Schrödinger equation. The ground state was computed by diagonalizing the field free Hamiltonian, while the continuum states were computed with a fourth-order Runge-Kutta algorithm and normalized with the Strömgren procedure \cite{seaton_determination_1962}. We then used these numerically exact states to compute the "exact" recombination dipole, and extract its zero for different values of $R$. This is shown with a solid green line on Fig. \ref{fig_minimum_position}. As predicted by the 3-step model, the zeros of the $exact$ recombination dipole perfectly match the position of the minima observed in the HHG spectra, as extracted from the long trajectory contributions. Note that, for small $R\lesssim1.6$ a.u., the mismatch is related to the difficulty of finding the minimum in the TDSE HHG spectrum, because, as can be seen in Fig. \ref{fig_TDSE_ampl_phase_long_R}, it is located too close to the cutoff. However, if we compute the recombination dipole using the "exact" continuum states and the LCAO ground state (dashed blue line), the position of the minimum is clearly underestimated, by at least 10 harmonic orders. On the contrary, if we compute the recombination dipole with the "exact" ground state and PW for the continuum (dotted pink and purple lines), the position of the minimum is now greatly overestimated, whether we use the $I_\mathrm{p}$ shift correction or not. Finally, when we perform both the LCAO and the PW approximations (dash-dotted orange and yellow lines), we see that the errors from both approximations compensate each other, providing a relatively good, but somehow accidental, agreement with the TDSE results. We note that this error compensation leads to a better agreement with the $I_\mathrm{p}$ shift correction. We thus want to draw attention that one always has to remain careful when considering model with several approximations. These error compensations can actually happen more often than we think \cite{labeye_tunnel_2018}.

We now turn to the \emph{shape} of the minimum and of the corresponding phase jumps. As already observed in \cite{lein_role_2002,lein_interference_2002,van_der_zwan_two-center_2010}, the  phase-jumps are smoothed, covering several harmonics, with a magnitude lower than $\pi$ rad (Figs. \ref{fig_TDSE_ampl_phase_short_R}b and \ref{fig_TDSE_ampl_phase_long_R}b).
The interference condition exhibits more surprising features. While the phase-jump is smoothed and negative for all values of $R$ for short trajectories (Fig. \ref{fig_TDSE_ampl_phase_short_R}b), the situation is completely different for the long trajectories. It is smoothed and positive for small $R$ , then it varies steeply at a critical internuclear distance $R_c=1.57$ a.u. beyond which the behavior reverses into a smooth negative jump as for the short trajectories (Fig. \ref{fig_TDSE_ampl_phase_long_R}b). Moreover, the depth of the minimum in the amplitude is directly related to the shape of the phase-jump: the steeper the jump, the deeper the minimum.

\begin{figure}
\begin{center}
 \includegraphics[width=\columnwidth]{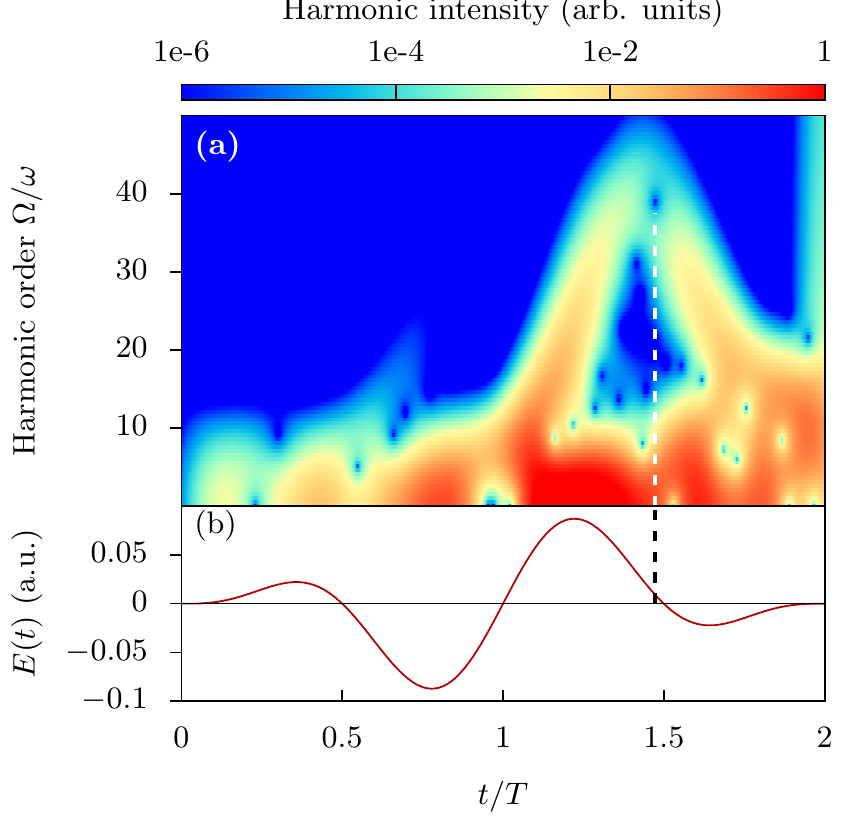}
  \caption{(Color online) Gabor transform of the ab initio dipole for the critical internuclear distance $R_c=1.57$ a.u. (a) and instantaneous value of the electric (b). The peak intensity is 2.65$\times10^{14}$ W.cm$^{-2}$. We marked with dashed-lines the  time associated with the minimum observed in the long trajectories.}
  \label{fig_gabor_Rc}
\end{center}
\end{figure}

Besides, we performed a time-frequency analysis of the harmonic dipole using Gabor transforms as in \cite{zhao_positioning_2012,spiewanowski_high-order-harmonic_2013,spiewanowski_field-induced_2014} to retrieve, for different values of $R$, the emission times of the harmonics for which the destructive interference occurs. Fig. \ref{fig_gabor_Rc} displays the Gabor transform of the dipole for the particular case of $R=R_c$. We found that the recollision time for the long trajectory associated with the interference frequency corresponds to an almost zero electric field ($E_{\mathrm{rec}}\simeq9.5\times10^{-3}$ a.u.). Moreover, greater (lower) values of $R$ lead to recombination in presence of  a negative (positive) electric field for the long trajectories while the field at recombination is always positive for the short trajectories.

\begin{figure}
\begin{center}
 \includegraphics[width=\columnwidth]{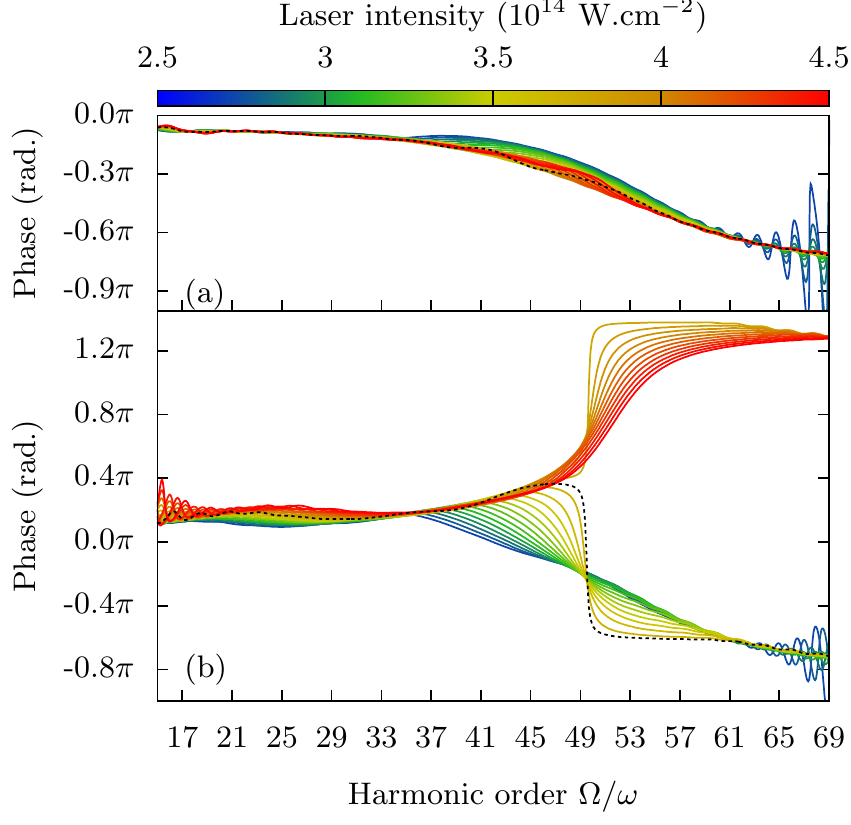}
  \caption{(Color online) Ab initio computations of the harmonic phase of a model H$_2$ relative to the atomic reference for a two-cycle laser pulse with peak intensities in the range $2.5-4.5\times10^{14}$ W.cm$^{-2}$ for short (a) and long trajectories (b). The dashed line corresponds to intensity $I_c=3.74\times10^{14}$ W.cm$^{-2}$.}
  \label{fig_TDSE_phase_S_L_I}
\end{center}
\end{figure}

To further examine the relation between the value of the electric field at the recombination and the shape of the phase-jump, we tuned the laser peak  intensity $I_\mathrm{L}$ from 2.5 to 4.5$\times10^{14}$ W.cm$^{-2}$ at fixed $R$. We chose $R = R_\mathrm{eq} = 1.4$ a.u. The corresponding trajectory-resolved phases are shown in Fig. \ref{fig_TDSE_phase_S_L_I}. As expected, the position of the destructive interference does not change with $I_\mathrm{L}$ and we observe a behavior for the phase-jumps which is similar to the one when varying $R$. It is smoothed and negative for all intensities at the short trajectories, while an equivalent smoothing and inversion behavior is present for the long trajectories. We found a critical intensity $I_c=3.74\times10^{14}$ W.cm$^{-2}$ at which the inversion occurs with a \textit{discontinuous} $\pm\pi$ rad phase-jump (see black-dashed curve in Fig. \ref{fig_TDSE_phase_S_L_I}b).

Using time-frequency analysis once again, this change of behavior is \textit{directly} related to the change of sign of the instantaneous electric field at recombination time, which is almost zero at $\Omega_0$ for the intensity $I_c$. 
We find that, in this case, the corresponding electric field amounts to $E_{\mathrm{rec}}=7.4\times10^{-3}$ a.u. This corresponds to $\equiv I_{\mathrm{rec}}=1.92\times10^{12}$ W.cm$^{-2}$, that is $\sim0.5\%$ of $I_c$.

\subsection{Molecular SFA}\label{subsec_results_SFA_mol}

\begin{figure}
\begin{center}
 \includegraphics[width=\columnwidth]{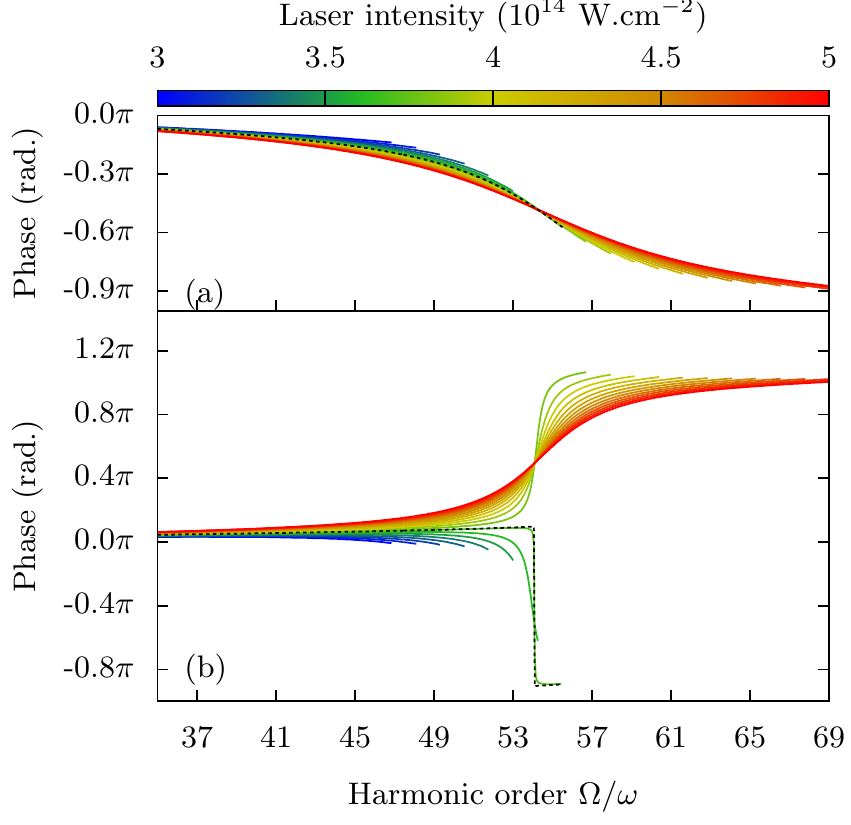}
  \caption{(Color online) Molecular SFA computations of the harmonic phase of a model H$_2$ normalized by the atomic SFA for a two-cycle laser pulse with peak intensities in the range $3-5\times10^{14}$ W.cm$^{-2}$ for short (a) and long trajectories (b). The dashed line corresponds to the intensity $I'_c=3.704\times10^{14}$ W.cm$^{-2}$.}
  \label{fig_SFA_mod_phase_S_L_I}
\end{center}
\end{figure}

To get more physical insights, the same study has been carried out within the molecular SFA derived in Sec. \ref{subsec_methods_SFA_mol}, following \cite{faria_high-order_2007}. We report in Fig. \ref{fig_SFA_mod_phase_S_L_I} the phase of the short and long trajectories as a function of the laser intensity. Again this phase is relative to the phase of atomic SFA presented in Sec. \ref{subsec_methods_SFA}, following \cite{lewenstein_theory_1994}. The separation of the short and long trajectory contributions in SFA is straightforward, being inherent to the search of saddle-point solutions. Consequently, no residual oscillation is encountered in contrast to the TDSE results. Nevertheless, the SFA model prevents the computation of harmonics above the cutoff, which limits the observation of the entire phase-jumps for the lowest intensities.

We observe that the behavior of the phase is very similar to that of the TDSE simulations. The major difference is the position of the destructive interference due to the LCAO and PW approximations in SFA, as mentioned before. Hence the destructive interference is observed exactly at the frequency given by Eq. (\ref{eq_omega0}). Furthermore, the intensity for which the phase is discontinuous and jumps by $\pm\pi$ rad is shifted to $I'_c=3.704\times10^{14}$ W.cm$^{-2}$. 

The SFA computations allowed us to obtain the precise value of the electric field {\it at recombination time}, for the long trajectories and for the inversion intensity $I'_c$. It is found to be $E_{\mathrm{rec}}=0.0034$ a.u., which is comparable to the one observed in the ab initio computations. The discrepancies are attributed to the difference between ab initio and SFA trajectory times \cite{risoud_quantitative_2013}.

Finally, we have performed simulations at the laser intensity, for which the electric field is \textit{exactly} zero at recombination time -  $I_z=3.641\times 10^{14}$ W.cm$^{-2}$, that is slightly lower than $I'_c$, with a difference of only $6\times 10^{12}$ W.cm$^{-2}$, i.e. $< 2\%$ -  and found that the jump \textit{is not} discontinuous and lower than $-\pi$ rad.

\subsection{Quantum Path Interferences}\label{subsec_QPI}

We have seen in the last two sections that the HHG phase behavior depends strongly on the type of trajectory (short vs long) and on the value of the electric field at recombination time, thus entangling the structural information contained in the two-centre interferences and the electron dynamics.

\begin{figure}[h]
\begin{center} 
  \includegraphics[width=\columnwidth]{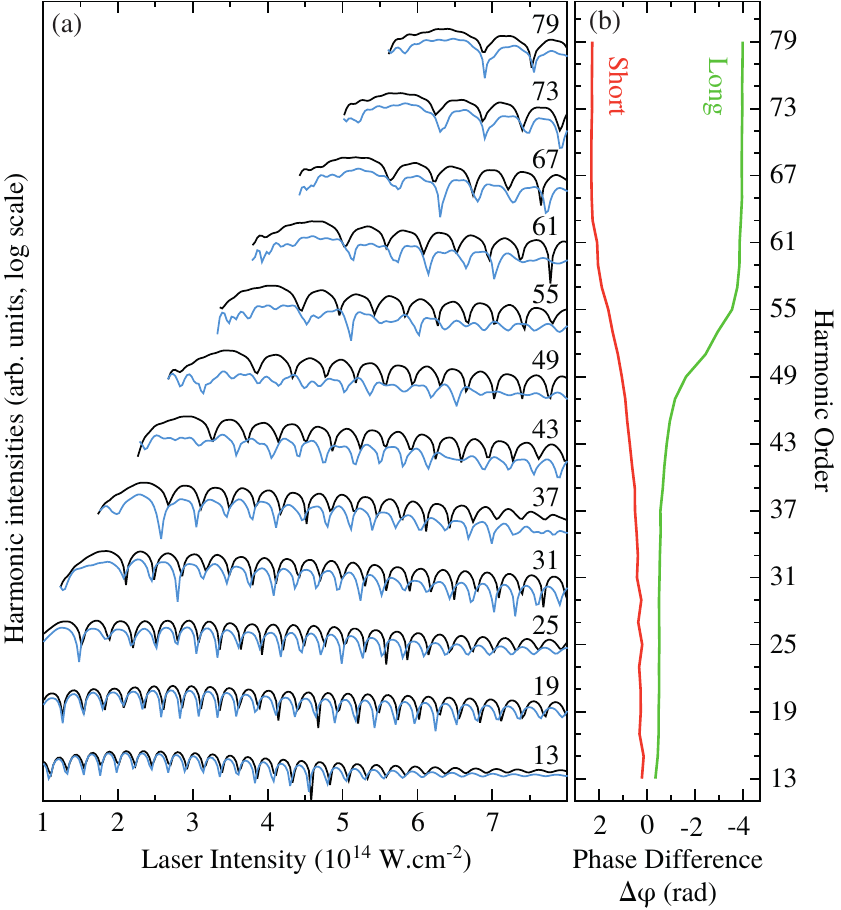}
  \caption{(Color online) (a) QPI for the model molecule with $R=1.4$ a.u. (blue lines) and the atomic reference (black lines) from TDSE computations. The curves corresponding to the different harmonic orders are shifted vertically for clarity; (b) phase difference $\Delta\varphi$ between the molecular and the atomic $I_\mathrm{L}\rightarrow\alpha$ Fourier transform components of the QPI, associated with the short (red lines) and long (green lines) trajectory contributions.}
  \label{Dbl_qpi}
 \end{center}
\end{figure}

In practice, the influence of the electron dynamics on a structural interference could be observed by following the phase difference between the short/long trajectory contributions when varying the laser intensity, e.g., by recording the Quantum Path Interferences (QPI) in the total harmonic dipole \cite{zair_quantum_2008,zair_molecular_2013}. We performed such an analysis in our TDSE simulations and show the results in Fig.~\ref{Dbl_qpi}(a), for both the molecule and the reference atom.
For each harmonic, we observe oscillations corresponding to the interferences between the short and long trajectories. They mainly come from the phases accumulated by the electron during its excursion in the continuum, i.e during the second step in HHG, which naturally depend on the trajectory duration. However, for molecules, one also has to take into account the phase of $\widetilde{d}_{\mathrm{rec}}$, which, as seen above, significantly depends on the continuum electron dynamics and is different for each contribution. In fact, as we can see in Fig.~\ref{fig_TDSE_phase_S_L_I}, for harmonics far from the interference (H49), the recombination dipole phase difference for the two trajectory contributions is either $\approx 0$ or $\approx2\pi$ rad, leading to oscillations in phase with the reference atom. For harmonics around H49, the phase difference is $\approx \pi$ rad and the QPI exhibit oscillations out of phase with the reference atom. This clear signature makes the QPI a method of choice to observe a pure structural interference modified by electron dynamics.

We extracted these phases by performing a $I_\mathrm{L}\rightarrow\alpha$ Fourier transform of the QPI, as described in \cite{balcou_quantum-path_1999}. We present in Fig.~\ref{Dbl_qpi}(b) the phase difference $\Delta\varphi$, between the model molecule and the reference atom. We observe phase-jumps that are an average of the different phase-jumps over the laser intensity, and that clearly reproduce the expected behavior; i.e.  smooth $\pi$ jumps occuring around H49, in opposite directions for the long and short trajectories respectively

Below, we shall show how the Taylor expansions of the molecular SFA times, action and dipole presented in Sec. \ref{subsec_methods_SFA_dl} give a comprehensive understanding of the variations of the HHG dipole phase jump with respect to $R$ and $I_L$.

\section{Interpretation}\label{sec_interpretation}

In this section we  examine the dipole $\widetilde{D}(\Omega)$ given by Eq. (\ref{eq_dipole_mol_dl}), investigating each of its terms (see Sec. \ref{subsubsec_total}). We  particularly focus on the role of the modified recombination dipole matrix element $\widetilde{d}_{\mathrm{rec}}$ of Eq. (\ref{eq_drec_tilde}).

\subsection{Recombination dipole matrix element}

\subsubsection{Explanation of the phase-jumps}

\begin{figure}
\begin{center}
 \includegraphics[width=\columnwidth]{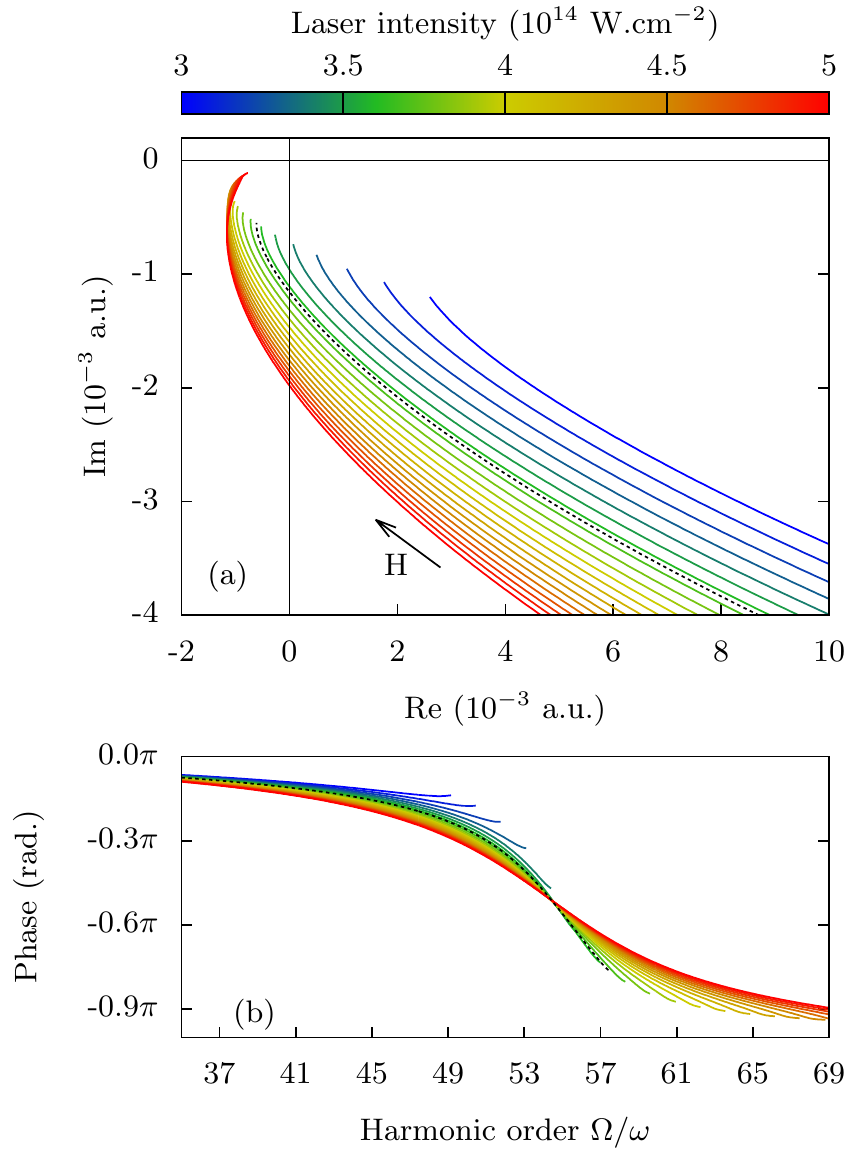}
  \caption{(Color online) Modified recombination dipole matrix element $\widetilde{d}_{\mathrm{rec}}$ (Eq. \ref{eq_drec_tilde}) in the complex plane for the short trajectories (a). The different paths correspond to different laser intensities between 3 and $5\times10^{14}$ W.cm$^{-2}$ and for the specific intensity of $I_z=3.641\times 10^{14}$ W.cm$^{-2}$ (black dashed line). Each path shows the evolution of $\widetilde{d}_{\mathrm{rec}}$ with the harmonic order H (see black arrow). Note that the position of the destructive interference corresponds approximately to the intersection between the curves and the imaginary axis (i.e. zero real part). The phase of $\widetilde{d}_{\mathrm{rec}}$ is reported in (b).}
  \label{fig_complex_drec_paths_S}
\end{center}
\end{figure}

\begin{figure}
\begin{center}
 \includegraphics[width=\columnwidth]{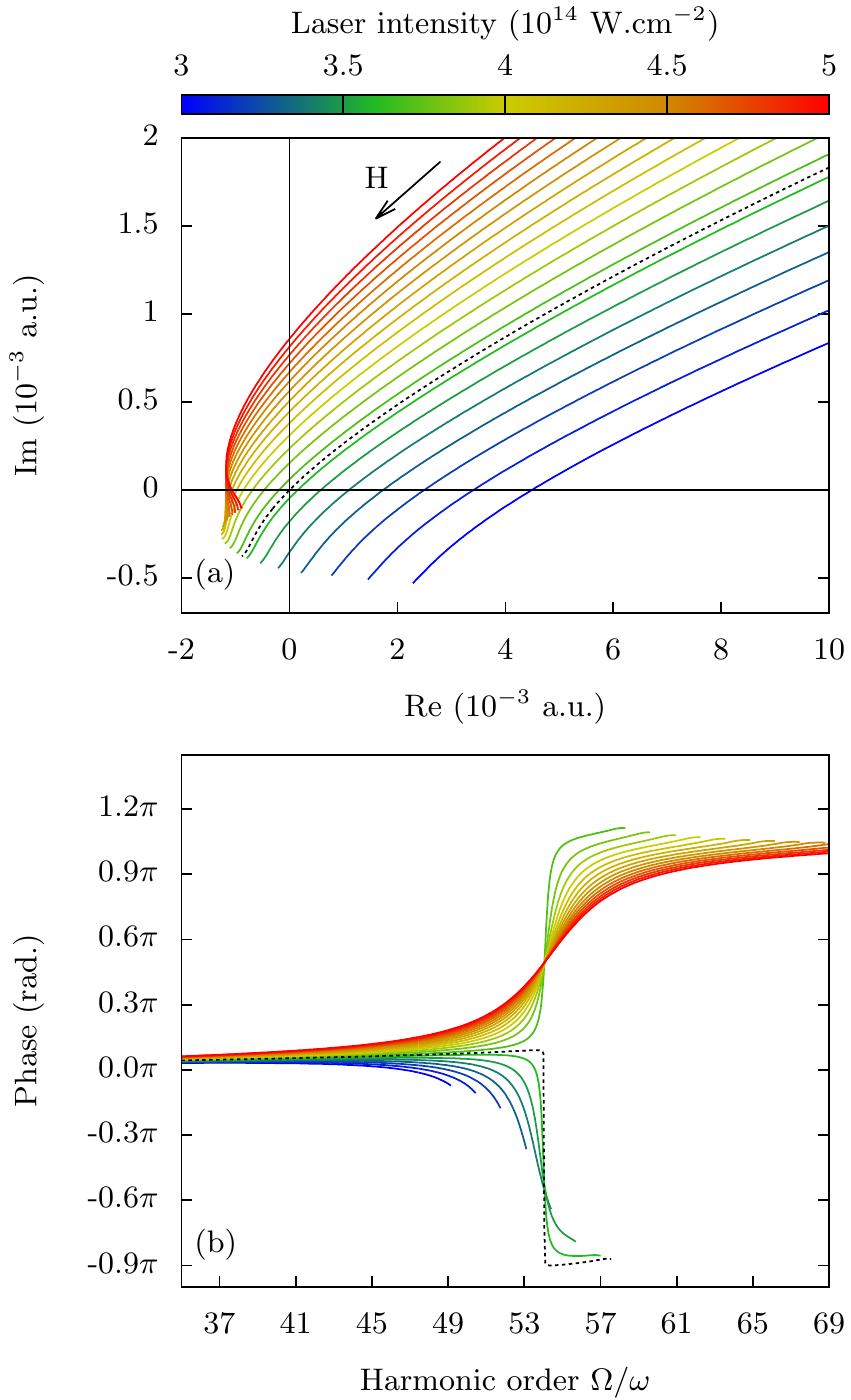}
  \caption{(Color online) Same as in Fig \ref{fig_complex_drec_paths_S} but for the long trajectories. The particular path which crosses the origin corresponds to an intensity of $I_z=3.641\times 10^{14}$ W.cm$^{-2}$ (black dashed line). The phase of $\widetilde{d}_{\mathrm{rec}}$ is reported in (b). 
  }
  \label{fig_complex_drec_paths_L}
\end{center}
\end{figure}

We already explained in Sec. \ref{sec_results} that the zero of $d_{\mathrm{rec}}$ (Eq. \ref{eq_d_rec}) encountered at $p_0$, see  Eq.(\ref{eq_omega0}), encodes the position of the destructive interference. For symmetry reasons (the ground state wave function is even), this recombination dipole matrix element is a real number in the PW approximation. Thus, it is constrained to the real axis and its evolution around $p_0$ from a positive to a negative value implies a discontinuity of its phase, with a jump of \textit{exactly} $-\pi$ rad.

However, the modified recombination dipole matrix element $\widetilde{d}_{\mathrm{rec}}$ expressed in Eq.~(\ref{eq_drec_tilde}) evolves in a totally different manner around $p_0$. As it contains a non zero imaginary part, it is not anymore constrained to the real axis and describes a path in the complex plane when varying the harmonic order. It turns out that this imaginary part is proportional to the instantaneous electric field at the recombination time, see Eq. (\ref{eq_R1}) and (\ref{eq_drec_tilde}). Fig. \ref{fig_complex_drec_paths_S}a displays the values of $\widetilde{d}_{\mathrm{rec}}$ in the complex plane for the short trajectories, for different laser intensities, when scanning the harmonic order. The path in the complex plane, followed at a fixed intensity, crosses the Re$(\widetilde{d}_{\mathrm{rec}})=0$ line for a \textit{non zero} imaginary part, implying a smooth variation of the phase. In addition, we observe that the higher the electric field at recombination, the further the path from the origin, and consequently, the smoother the phase-jump. 
Furthermore, as the electric field at the recombination is always positive for the short trajectories, as pointed out in Sec \ref{sec_results}, we observe a systematic negative phase-jump as shown in Fig. \ref{fig_complex_drec_paths_S}b. This behaviour is fully imprinted on the phase of the full HHG dipole as seen in Fig.~\ref{fig_SFA_mod_phase_S_L_I}a. Note that  for the lowest intensities the phase-jump is not complete due to the low cutoff value.

The situation is completely different for the long trajectories for which the recombination times are distributed across two laser half cycles: the electric field is first positive, crosses zero and is then negative when decreasing the harmonic order. As we can see in Fig. \ref{fig_complex_drec_paths_L}a for the particular intensity denoted $I_{z}$ (black dashed line), the path followed by $\widetilde{d}_{\mathrm{rec}}$ crosses the origin. This corresponds to a zero electric field at recombination time and thus a zero imaginary part, when the real part also cancels. Therefore, we retrieve the ideal case for which the phase is discontinuous and jumps to $\pm\pi$ rad (see Fig. \ref{fig_complex_drec_paths_L}b). For the other intensities, the laser field is not zero and the phase evolves slowly, varying from zero to $-\pi$ ($\pi$) rad for a positive (negative) electric field at the recombination.


Thus, we fully understand the behavior of the recombination dipole matrix element $\widetilde{d}_{\mathrm{rec}}$, which mainly drives the HHG phase evolution along the 2-center destructive interference. Its deviations from the ideal case are governed by the continuum dynamics of the electron during the second step of HHG. This leads to  a phase-jump whose shape depends on the instantaneous value of the electric field at recombination time. Since the electrons following short or long trajectories recombine with the molecule within different time windows, the corresponding phase-jumps behave completely differently. As we explained, three different behaviors are notably observed for the long trajectories. We analyze below quantitatively the conditions for which we can observe these three situations.


\subsubsection{Inversion condition}

In this section, we quantify the conditions to observe the inversion in the phase-jump for the long trajectories in our simulations either as a function of internuclear distance $R$ (see Fig. \ref{fig_TDSE_ampl_phase_long_R} in Sec. \ref{subsec_results_ab_initio}) or as a function of the laser intensity $I_\mathrm{L}$ (see Fig. \ref{fig_TDSE_phase_S_L_I} in Sec. \ref{subsec_results_ab_initio} or Fig. \ref{fig_SFA_mod_phase_S_L_I} in Sec. \ref{subsec_results_SFA_mol}).

For the long trajectory contributions, the electric field changes sign at a single harmonic frequency denoted $\Omega_{\mathrm{null}}$. Hence, harmonics below the cutoff and beyond (below) $\Omega_{\mathrm{null}}$ are emitted in presence of a positive (negative) electric field.

The value of $\Omega_{\mathrm{null}}$ can be obtained with atomic classical calculations  \cite{corkum_plasma_1993}. We thus solved the Newton equation for an electron freed at ionization time $t'$ and brought back to its initial position $x=0$ at recombination time $t$ under the influence of the laser field only.
The recollision time for which the laser field is zero is $t_c=3\pi/2\omega$ (see Eq. \ref{eq_laser_SFA}), leading to the corresponding ionization time:
\begin{equation}
  t'_c\simeq\frac{1}{\big(\frac{3\pi}{2}-\frac{2}{3\pi}\big)\omega}\,.
\end{equation}
Thus, the corresponding critical harmonic frequency is:
\begin{equation}\label{eq_omega_turn}
  \Omega_{\mathrm{null}} = I_\mathrm{p}+\frac{E_\mathrm{L}^2}{2\omega^2}[\sin(\omega t'_c)+1]^2 \approx I_\mathrm{p}+2.98 U_p\,,
\end{equation}
where $U_p=E_\mathrm{L}^2/4\omega^2=I_\mathrm{L}/4\omega^2$ is the ponderomotive energy.

Therefore, we have two knobs to tune the relative values of $\Omega_0^{(\mathrm{PW})}$, Eq. (\ref{eq_omega0}), and $\Omega_{\mathrm{null}}$, Eq. (\ref{eq_omega_turn}), i.e. the internuclear distance $R$ for the former and the laser intensity for the latter. By varying these knobs, we can put our system in the three cases that we previously discussed: i) $\Omega_0^{(\mathrm{PW})}<\Omega_{\mathrm{null}}$, the phase-jump for the long trajectories is smoothed and has an opposite direction than the one of the short trajectories; ii) $\Omega_0^{(\mathrm{PW})}=\Omega_{\mathrm{null}}$, the phase-jump for the long trajectories is discontinuous and equals to $\pm\pi$ rad; iii) $\Omega_0^{(\mathrm{PW})}>\Omega_{\mathrm{null}}$, the phase-jump for the long trajectories is smoothed and has the same direction as the one of the short trajectories.

Furthermore, we found that the equality of $\Omega_0^{(\mathrm{PW})}$ and $\Omega_{\mathrm{null}}$ can be obtained for an infinity of couples $(I_\mathrm{L},R)$, approximately verifying the relation:
\begin{equation}\label{eq_Rx_f_Il}
  R\simeq2.58\frac{\omega}{\sqrt{I_\mathrm{L}}}\,.
\end{equation}

Finally, we would like to note that, except in our ab initio computations, the entire phase-jump is not observed because $\Omega_{\mathrm{null}}$ is close to the cutoff frequency, i.e.  $I_\mathrm{p}+3.17 U_p$ \cite{lhuillier_high-order_1993}, as we can see in Fig. \ref{fig_complex_drec_paths_L}a or Fig. \ref{fig_complex_drec_paths_L}b. Unfortunately one cannot expand the region between these two frequencies by changing the laser intensity since both depend linearly on it.

The SFA-based analysis of $\widetilde{d}_{\mathrm{rec}}$ thus allows us to relate the discontinuous phase to the recombination in presence of a zero electric field. However, in Sec. \ref{sec_results}, both for TDSE and molecular SFA computations, we observed that the sharpest phase-jump is obtained when the electric field at recombination time is {\it not exactly} zero. In the following section we derive the first order expansion of the prefactor $C_{\alpha\beta}$ and explain the origin of this shift.

\subsection{Prefactor correction}\label{subsec_interp_extra_phase}

When we obtained the first order expression of the SFA molecular dipole in Sec. \ref{subsubsec_total} we neglected the first order of the molecular prefactor $C_{\alpha\beta}$. This first approach allowed us to get a factorized expression for the molecular dipole (\ref{eq_dipole_mol_dl}) that led us to important physical insights on the inversion of the $\pi$ jump observed in the HHG phase associated with the long trajectories. 
However, as mentionned above, discrepancies remain on the exact values of the electric field for which we observe the phase jump inversion. To assess the origin of this difference and to go deeper in the analysis we complete here the analytic derivation of the expression of $C_{\alpha\beta}$ up to the first order as

\begin{align}
\label{eq_prefactor_first_order}
C_{\alpha\beta}(t_{\alpha\beta},t'_{\alpha\beta})=&C_\mathrm{at}(t_\mathrm{at},t'_\mathrm{at})\Bigg\lbrace 1 + \frac{R}{2} \Bigg[  \frac{(-1)^\alpha E(t'_\mathrm{at})}{2\left[p_\mathrm{at}+A(t'_\mathrm{at})\right]^2} \nonumber\\
&+\frac{(-1)^\beta E(t_\mathrm{at})}{2\left[p_\mathrm{at}+A(t_\mathrm{at})\right]^2} \nonumber\\
&+\frac{(-1)^\alpha}{(t_\mathrm{at}-t'_\mathrm{at})\left[p_\mathrm{at}+A(t'_\mathrm{at})\right]}\\
&-\frac{(-1)^\beta}{(t_\mathrm{at}-t'_\mathrm{at})\left[p_\mathrm{at}+A(t_\mathrm{at})\right]}\Bigg] \Bigg\rbrace\nonumber.
\end{align}
For details, see Appendix B.

We then need to sum over the four trajectories (i.e. over $\alpha$ and $\beta$) to obtain the full first order expression of the total dipole (\ref{eq_tot_dip_mod_sdp}). The latter can again be factorized in the same form as in the atomic case (\ref{eq_dipole_SFA}):
\begin{align}\label{eq_dipole_mol_dl_corrected}
  D(\Omega)  = &C(t_\mathrm{at},t'_\mathrm{at})\widehat{d}_{\mathrm{rec}}(p_\mathrm{at}+A(t_\mathrm{at}),t_\mathrm{at},t'_\mathrm{at}) \nonumber \\
                       & \times \widehat{d}_{\mathrm{ion}}(p_\mathrm{at}+A(t'_\mathrm{at}),t_\mathrm{at},t'_\mathrm{at}) e^{-iS_\mathrm{at}}  + O(R^2)\,,
\end{align}
where the \emph{modified} recombination $\widehat{d}_\mathrm{rec}$ and ionization $\widehat{d}_\mathrm{ion}$ dipoles now include contributions from the prefactor expansion. Their expressions read:
\begin{multline}\label{eq_drec_hat}
  \widehat{d}_{\mathrm{rec}}(p,t,t')=2\mathcal{R}(p)\cos\left(\frac{pR}{2}\right) \\
  +2i\,\mathcal{Q}(p,t,t')R\sin\left(\frac{pR}{2}\right)\,,
\end{multline}
with 
\begin{align}
\label{eq_Q_def}
  \mathcal{Q}(p,t,t')=&-\frac{E(t)}{4}\frac{\partial\bar{\phi}_s}{\partial p}(p)-\frac{E(t)}{8p}\bar{\phi}_s(p) -\frac{\bar{\phi}_s(p)}{4\left(t-t'\right)}
\end{align}
and
\begin{multline}\label{eq_dion_hat}
  \widehat{d}_{\mathrm{ion}}(p,t,t')=2\mathcal{I}_\mathrm{at}(p,t')\cos\left(\frac{pR}{2}\right) \\
  +2i\,\mathcal{K}(p,t,t')R\sin\left(\frac{pR}{2}\right)\,,
\end{multline}
with
\begin{align}
\mathcal{K}(p,t,t')=&\frac{E(t')}{4}\bar{\phi}_s(p) -i\frac{E^2(t')}{4p}\frac{\partial^2 \bar{\phi}_s}{\partial p^2}(p)\nonumber\\
&+i{\pdv{\bar{\phi}_s}{p}} (p) \Bigg[ \frac{3}{2(t-t')p}-\frac{E(t')}{4I_\mathrm{p}} +\frac{\omega^2A(t')}{p} \Bigg]\,.
\end{align}
We thus recover the same kind of expressions as before, with additional terms in the \emph{modified} recombination $\widehat{d}_\mathrm{rec}$ and ionization $\widehat{d}_\mathrm{ion}$ dipoles that accounts for the influence of the prefactor. In fact, these expressions are reminiscent of Eqs. (\ref{eq_R1}) and (\ref{eq_I1}) where $\mathcal{R}^{(1)}$ and $\mathcal{I}^{(1)}$ defined in Eqs. (\ref{eq_drec_tilde}) and (\ref{eq_dion_tilde}) are now replaced by $\mathcal{K}$ and $\mathcal{Q}$, respectively. Note that these extra terms correspond to the $\zeta$ parameter introduced in Eq. (4) of Ref. \cite{risoud_laser-induced_2017}. When looking at the expression of $\widehat{d}_\mathrm{rec}$, we remark that, because of this new terms, the imaginary part of $\widehat{d}_\mathrm{rec}$, which is related to the real part of $\mathcal{Q}$ in Eq (\ref{eq_Q_def}), is no longer proportional to the electric field at recombination time. It thus  explains why the sharp phase jump occurs for a small \emph{non zero} electric field at recombination time.

\begin{figure}
\begin{center}
 \includegraphics[width=\columnwidth]{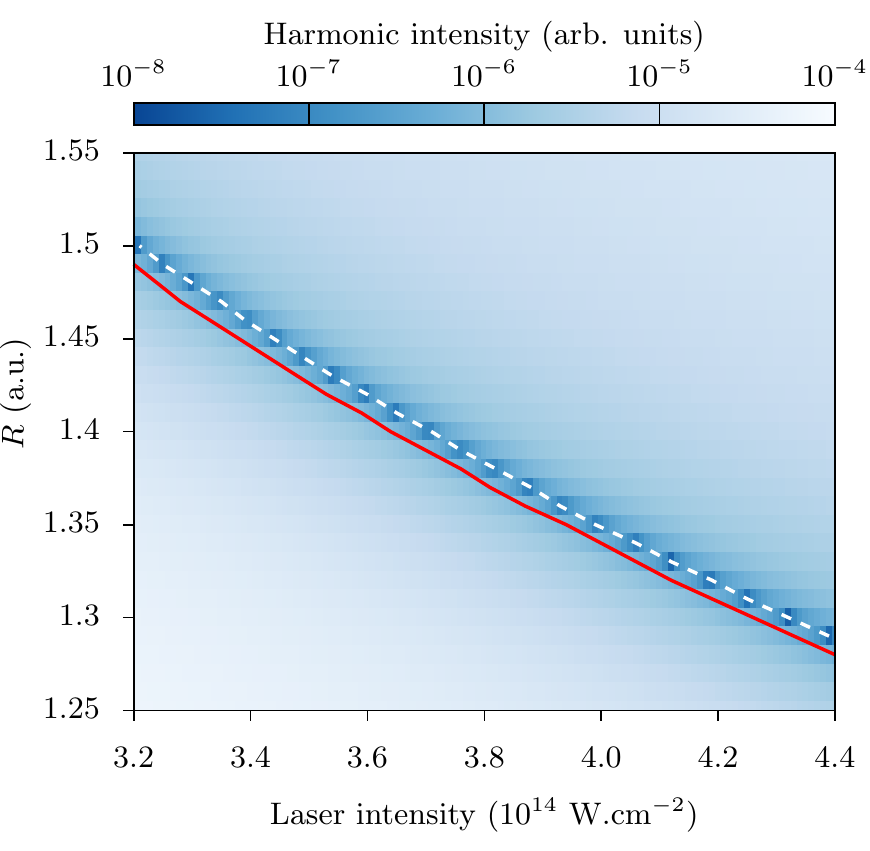}
  \caption{(Color online) Value of the minimum of the intensity of the long trajectory contribution in the HHG spectrum as a function of the laser intensity and nuclear distance $R$. The colormap corresponds to the value of the minimum of the molecular SFA dipole Eq. (\ref{eq_tot_dip_mod_sdp}). The lines corresponds to the couples $(R,I_\mathrm{L})$ for which the modified recombination dipole crosses zero in the complex plane, solid red line $\widetilde{d}_\mathrm{rec}$ Eq. (\ref{eq_drec_tilde}), dotted white line $\widehat{d}_\mathrm{rec}$ Eq. (\ref{eq_drec_hat}).}
  \label{fig_minima}
\end{center}
\end{figure}

To  better understand the effect of these new terms in the modified recombination dipole, we compare both $\widetilde{d}_\mathrm{rec}$ and $\widehat{d}_\mathrm{rec}$ to the full SFA results in Fig. \ref{fig_minima}. The color map indicates the value of the minimum in the full SFA dipole Eq. (\ref{eq_tot_dip_mod_sdp}). The darkest pixels correspond to the couples $(R,I_\mathrm{L})$ for which the minimum is the deepest and the phase jump is the sharpest. The lines show the couples $(R,I_\mathrm{L})$ for which the modified recombination dipole crosses the origin of the complex plane, the solid red line is for $\widetilde{d}_\mathrm{rec}$ (without the prefactor contribution) and the dotted white line is for $\widehat{d}_\mathrm{rec}$ (with the prefactor contribution). It is striking that the prefactor contribution allows to completely recover the correct behavior. Indeed, the zeros of $\widehat{d}_\mathrm{rec}$ perfectly match the deep minima in the SFA spectra. As a consequence, the study of the modified recombination dipole $\widehat{d}_\mathrm{rec}$ is sufficient to fully recover the \emph{position} and the \emph{shape} of both the structural minimum and the phase jump expected in diatomic molecular HHG.

\section{Conclusion}\label{sec_conclusion}

In this paper, we have revisited the phenomenon of two-center interferences in molecular HHG, with a strong focus on its spectral phase. Using both TDSE and molecular SFA computations, which display similar results, we have shown that the phase for the short and long trajectory contributions behave in totally different ways. In particular, the destructive interference phase-jump for the former is always smooth and negative while for the latter the features of the jump strongly depend on the internuclear distance and on the laser intensity.

With the help of a time-frequency analysis of the numerical results we have attributed this effect to the value (magnitude and sign) of the driving electric field when the harmonics are emitted, i.e at recombination time. We confirmed these findings analytically by performing a Taylor expansion of the molecular SFA around the reference atomic SFA. As was done in \cite{spiewanowski_high-order-harmonic_2013,spiewanowski_field-induced_2014}, this expansion allowed us to factorize the SFA dipole expression with a \emph{modified} recombination dipole matrix element. We found that if we neglect the molecular specificities of the prefactor involved in the factorized expression, then this modified recombination dipole behaves very similarly to the one found in  \cite{spiewanowski_high-order-harmonic_2013,spiewanowski_field-induced_2014}: it exhibits a sharp phase jump if and only if the electric field vanishes at the recombination time. However, we showed that this is not in perfect agreement with our TDSE simulations, for which the sharp phase jump appears for a small \emph{non zero} electric field at recombination. By consistently taking into account the contribution from the prefactor, our approach allows to fully explain this behavior.

Our study also confirms that structural two-center interferences can be strongly affected by the instantaneous value  of the laser field at recombination and that, in this case, factorization of the HHG dipole proposed in the Quantitative Rescattering Theory \cite{le_theory_2008,le_quantitative_2009,le_quantitative_2013} is not straightforward. More precisely, when one deals explicitly with the two atomic centers of the molecule, this leads to four different classes of electron trajectories, which imprint their signatures inside the recombination dipole matrix element. Thus, the freed electron dynamics and the recombination steps are entangled and the transition dipole matrix elements cannot be assimilated to the one of the static case.

Finally, as the phases of the short and long trajectories evolve differently, an interferometric experimental setup based on Quantum Path Interferences \cite{zair_quantum_2008,zair_molecular_2013} should be able to reveal their signatures.

\section*{Acknowledgments}

We acknowledge financial support from the LABEX Plas@Par-ANR-11-IDEX-0004-02 and ANR-10-LABX-0039-PALM, the programs ANR11-EQPX0005-ATTOLAB, ANR-15-CE30-0001-01-CIMBAAD and the European Networks ITN-ATTOFEL and ITN-MEDEA. CL acknowledges financial support from the Wiener Wissenschafts- und TechnologieFonds (WWTF) project No MA16-066 ("SEQUEX")

\appendix
\section{Second order corrections}
From Eqs. (\ref{eq_SP_mol}) and \ref{eq_sol_mol_DL}, we obtain the following set of equations for the second order terms:
\begin{subequations}
\label{eq_SP_mol_2nd}
\begin{align}
\label{eq_SP_mol_2nd_p}
0&=\left[p_\mathrm{at}+ A(t_\mathrm{at})\right]\Delta t^{(2)}_{\alpha\beta} - \left[p_\mathrm{at}+ A(t'_\mathrm{at})\right]\Delta t'^{(2)}_{\alpha\beta}\nonumber\\ 
& + (t_\mathrm{at}-t'_\mathrm{at})\Delta p^{(2)}_{\alpha\beta} 
    - \frac{E(t_\mathrm{at})R^2}{8\left[p_\mathrm{at}+ A(t_\mathrm{at})\right]} - \frac{E(t'_\mathrm{at})R^2}{8\left[p_\mathrm{at}+ A(t'_\mathrm{at})\right]}, \\
\label{eq_SP_mol_2nd_t}
0&=-\left[p_\mathrm{at}+ A(t_\mathrm{at})\right]E(t_\mathrm{at})\Delta t^{(2)}_{\alpha\beta} + \left[p_\mathrm{at}+ A(t_\mathrm{at})\right]\Delta p^{(2)}_{\alpha\beta} \nonumber \\
   &+ \frac{E(t_\mathrm{at})^2R^2}{16(\Omega-I_\mathrm{p})}+\frac{\omega^2A(t_\mathrm{at})R^2}{8\left[p_\mathrm{at}+ A(t_\mathrm{at})\right]}, \\
\label{eq_SP_mol_2nd_tp}
0&=-\left[p_\mathrm{at}+ A(t'_\mathrm{at})\right]E(t'_\mathrm{at})\Delta t'^{(2)}_{\alpha\beta} + \left[p_\mathrm{at}+ A(t'_\mathrm{at})\right]\Delta p^{(2)}_{\alpha\beta} \nonumber \\
   &- \frac{E(t'_\mathrm{at})^2R^2}{16I_\mathrm{p}}+\frac{\omega^2A(t'_\mathrm{at})R^2}{8\left[p_\mathrm{at}+ A(t'_\mathrm{at})\right]}. \\ 
\end{align}
\end{subequations}
That can be inverted to get the second order terms:
\begin{subequations}
\label{eq_SP_mol_2nd_sol}
\begin{align}
\label{eq_SP_mol_2nd_sol_p}
\Delta p^{(2)}_{\alpha\beta}&=\frac{\omega^2R^2}{8\Delta}\left(\frac{A(t'_\mathrm{at})E(t_\mathrm{at})}{p_\mathrm{at}+A(t'_\mathrm{at})}-\frac{A(t_\mathrm{at})E(t'_\mathrm{at})}{p_\mathrm{at}+A(t_\mathrm{at})}\right), \\
\label{eq_SP_mol_2nd_sol_t}
\Delta t^{(2)}_{\alpha\beta}&=\frac{E(t_\mathrm{at})R^2}{16(\Omega-I_\mathrm{p})\left[p_\mathrm{at}+A(t_\mathrm{at})\right]}+\frac{\omega^2A(t'_\mathrm{at})R^2}{8\left[p_\mathrm{at}+A(t'_\mathrm{at})\right]\Delta} \nonumber\\
&+\frac{\omega^2A(t_\mathrm{at})R^2}{16(\Omega-I_\mathrm{p})\Delta}\left[(t_\mathrm{at}-t'_\mathrm{at})E(t'_\mathrm{at})+p_\mathrm{at}+A(t'_\mathrm{at})\right], \\
\label{eq_SP_mol_2nd_sol_tp}
\Delta t'^{(2)}_{\alpha\beta}&=-\frac{E(t'_\mathrm{at})R^2}{16I_\mathrm{p}\left[p_\mathrm{at}+A(t'_\mathrm{at})\right]}-\frac{\omega^2A(t_\mathrm{at})R^2}{8\left[p_\mathrm{at}+A(t_\mathrm{at})\right]\Delta}\nonumber \\
&-\frac{\omega^2A(t'_\mathrm{at})R^2}{16I_\mathrm{p}\Delta}\left[(t_\mathrm{at}-t'_\mathrm{at})E(t_\mathrm{at})+p_\mathrm{at}+A(t_\mathrm{at})\right].
\end{align}
\end{subequations}

\section{Prefactor Derivation}
We first use the chain rule for computing the derivative of the composition of functions, to find
\begin{align}
{\pdv[2]{S^{(p)}_{\alpha\beta}}{t}} (t,t')=&-E(t)\cdot[p_{\alpha\beta}+A(t)]-\frac{[p_{\alpha\beta}+A(t)]^2}{t-t'}\nonumber\\
 &+\frac{(-1)^\beta R}{2}\omega^2A(t),\\
\frac{\partial^2S^{(p)}_{\alpha\beta}}{\partial t^{\prime 2}}(t,t')=&E(t')\cdot[p_{\alpha\beta}+A(t')]-\frac{[p_{\alpha\beta}+A(t')]^2}{t-t'}\nonumber\\
&-\frac{(-1)^\alpha R}{2}\omega^2A(t'),\\
{\pdv{S^{(p)}_{\alpha\beta}}{t}{t'}} (t,t')&=\frac{[p_{\alpha\beta}+A(t)]\cdot[p_{\alpha\beta}+A(t')]}{t-t'}.
\end{align}
We can thus expand the Hessian determinant as:
\begin{align}
\det S''_{\alpha\beta,p}(t_{\alpha\beta},t'_{\alpha\beta})=&\det S_p''(t_\mathrm{at},t'_\mathrm{at})\Bigg\lbrace1-\frac{R}{2}\Bigg[\frac{(-1)^\alpha E( t'_\mathrm{at} )}{ \left[p_{\mathrm{at}}+ A ( t'_\mathrm{at} )\right]^2 }\nonumber\\
& + \frac{(-1)^\beta E( t_\mathrm{at} )}{\left[ p_{\mathrm{at}}+ A ( t_\mathrm{at} )\right]^2 } \nonumber\\
&-\frac{(-1)^\alpha}{ (t_\mathrm{at}-t'_\mathrm{at})\left[p_{\mathrm{at}}+ A ( t'_\mathrm{at} )\right] } \nonumber\\
&+\frac{(-1)^\beta}{ (t_\mathrm{at}-t'_\mathrm{at})\left[p_{\mathrm{at}}+ A ( t_\mathrm{at} )\right] } 
\Bigg]\Bigg\rbrace \nonumber\\
&+O(R^2).
\end{align}
%


\end{document}